\documentclass[twocolumn,tighten,times,linenumbers]{aastex63}
\usepackage{sidecap}
\usepackage{amsmath,amstext}
\usepackage{mathtools}
\usepackage{afterpage}
\usepackage{lipsum}
\usepackage{capt-of}
\usepackage{relsize}
\usepackage{newtxtext,newtxmath}
\usepackage[all]{hypcap} 
\usepackage{verbatim}
\usepackage{graphicx}
\usepackage{footnote}
\usepackage{float}
\usepackage{amsmath}
\usepackage{enumitem}
\usepackage{amssymb}
\usepackage{array}
\usepackage{gensymb}
\usepackage[graphicx]{realboxes}
\usepackage[version=4]{mhchem}
\usepackage{graphicx,epsf}
\usepackage{subfigure}
\usepackage{units}
\usepackage{natbib}

\usepackage{rotating}
\usepackage{nicefrac}
\usepackage{xtab} 
\usepackage{soul}

\usepackage{savesym}
\savesymbol{tablenum}
\usepackage{siunitx}
\restoresymbol{SIX}{tablenum}

\def\msun{M$_{\odot}$}

\newcommand{\Msun}{{\rm M}_\odot}

\newcommand{\niVI}{${}^{56}\textrm{Ni}$}

\newcommand{\caIV}{${}^{40}\textrm{Ca}$}
\newcommand{\TiIV}{${}^{44}\textrm{Ti}$}
\newcommand{\CHE}{CHE Israel Excellence Fellowship}

\def\cas{{Ca-rich}}
\def\fca{{CaSNe}}
\def\cast{{Ca-rich}}
\def\flash{{\texttt{FLASH}}}
\let\ts=\thinspace

\def\one{\ts {\,\sc i}}
\def\two{\ts {\,\sc ii}}
\def\three{\ts {\,\sc iii}}
\def\four{\ts {\,\sc iv}}
\def\five{\ts {\sc v}}

\nolinenumbers

\graphicspath{{./}{figures/}}
\shorttitle{Ca-rich SNe from CO-WDs disruptions by HeCO-WDs}
\shortauthors{Zenati et al.}
\begin{document}

\title{The origins of Calcium-rich supernovae from disruptions of CO white-dwarfs by hybrid He-CO white-dwarfs}

\correspondingauthor{Yossef Zenati, Hagai B. Perets}
\email{yzenati1@jhu.edu, hperets@physics.technion.ac.il}

\author[0000-0002-0632-8897]{Yossef Zenati}
\altaffiliation{\CHE}
\affil{Physics and Astronomy Department, Johns Hopkins University, Baltimore, MD 21218, USA}
\affil{Technion - Israel Institute of Technology, Physics department, Haifa Israel 3200002}

\author[0000-0002-5004-199X]{Hagai B. Perets}
\affil{Technion - Israel Institute of Technology, Physics department, Haifa Israel 3200002}

\author{Luc Dessart}
\affil{Institut d’Astrophysique de Paris, CNRS-Sorbonne Université, 98 bis boulevard Arago, F-75014 Paris, France}

\author[0000-0002-3934-2644]{Wynn V. Jacobson-Galán}
\affil{Department of Astronomy and Astrophysics, University of California, Berkeley, CA 94720, USA}

\author[0000-0002-2998-7940]{Silvia Toonen}
\affil{Anton Pannekoek Institute for Astronomy, University of Amsterdam, 1090 GE Amsterdam, The Netherlands}

%\author[0000-0003-4768-7586]{R.~Margutti}
%\affil{Department of Astronomy and Astrophysics, University of California, Berkeley, CA 94720, USA}

\author[0000-0002-4410-5387]{Armin Rest}
\affil{Space Telescope Science Institute, 3700 San Martin Dr., Baltimore, MD 21218, USA}
\affil{Physics and Astronomy Department, Johns Hopkins University, Baltimore, MD 21218, USA}

\begin{abstract}

Calcium-rich explosions are very faint ($\rm M_{B}\sim-15.5$), type I supernovae (SNe) showing strong Ca-lines, mostly observed in old stellar environments. Several models for such SNe had been explored and debated, but non were able to consistently reproduce the observed properties of \cas{} SNe, nor their rates and host-galaxy distributions. Here we show that the disruptions of low-mass carbon-oxygen (CO) white dwarfs (WDs) by hybrid Helium-CO (HeCO) WDs during their merger could explain the origin and properties of such SNe. We make use of detailed multi-dimensional hydrodynamical-thermonuclear (FLASH) simulations to characterize such explosions. We find that the accretion of CO material onto a HeCO-WD heats its He-shell and eventually leads to its "weak" detonation and ejection and the production of a sub-energetic $\sim10^{49}$erg \cas{} SN, while leaving the CO core of the HeCO-WD intact as a hot remnant WD, possibly giving rise to \textit{X}-ray emission as it cools down. We model the detailed light-curve and spectra of such explosions to find an excellent agreement with observations of Ia/c \cas{} might potentially also Ib \cas{} SNe. We thereby provide a viable, consistent model for the origins of \cas{}. These findings can shed new light on the role of \cas{} in the chemical evolution of galaxies and the intra-cluster medium, and their contribution to the observed 511 kev signal in the Galaxy originating from positrons produced from \TiIV{} decay. Finally, the origins of such SNe points to the key-role of HeCO WDs as SN progenitors, and their potential role as progenitors of other thermonuclear SNe including normal Ia.
\end{abstract}


\section{Introduction}     \label{sec:intro}
The discovery of the type Ib (He-rich, hydrogen poor) SN 2005E in a remote position of an early type galaxy, where no evidence of star-forming regions were detected lead to the detailed study and characterization of a novel type of faint \cas{} SNe (hereafter denoted \fca{}). 

The ejected mass from such SNe was inferred to be low ($\rm \lesssim 0.6M_{\odot}$) and they are typically found in old stellar environments; in addition their photometric and spectroscopic properties are peculiar and differ from typical type Ib SNe. The light curve (lc) of those typical \cas{} are characterized by peak luminosity of -$14$ to -$16.5$ mag, with fast rise times $\rm t_r \lesssim 15$ days \citep{Perets+10,Kasliwal+12,WynnV+20b}. These SNe showed strong {Ca}{II} lines, leading to their original identification as \cas{} \footnote{Some studies suggested that these explosions do not in fact produce more Ca in abundance relative to O, and suggested to term such SNe as Ca-strong SNe, referring to the line strength. However, in the models shown here, a large fraction of the ejecta is composed of Ca, as originally suggested \citep{Perets+10}, and we therefore refer to these objects as \cas{} SNe, as originally suggested.}  

Many of these faint SNe showed evidence for Helium lines and were identified as type Ib SNe \citep{Perets+10}, although later surveys identified faint, \cas{} Ia/c SNe showing no clear He lines, suggesting a likely continuum of spectroscopic properties showing a range of events, from SN Ia-like features (Ca-Ia objects) to those with SN Ib/c-like features (Ca-Ib/c objects) at peak light \citep{De+20}.

Detection of such SNe is difficult, given their low-luminosity and rapid evolution. However, the inferred rates of such SNe are likely in the range of 5-20$\%$ of the normal type Ia SNe rates \citep{Perets+10,Kasliwal+12, De+20}. The demographics of their host galaxies differ from that of normal type Ia SNe. Normal type Ia SNe are inferred to have a delay time distribution favoring short delays, most of which exploding up to a Gyr following their stellar progenitors formation \citep{Maoz+18}. In contrast,  the majority of \cas{} SNe were observed in early type galaxies, and are generally observed in old environments, far from star-forming regions \citep{Perets+10,Perets+11,Kasliwal+12,Lym+13,De+20}, suggesting a much more extended delay time distribution. \cas{} SNe also appear to have a relatively large offsets from their host galaxy nuclei \citep{Perets+10,Kasliwal+10,Foley+15,De+20}, however, this was shown to be the result of their high frequency in early type galaxies, which stellar halos are large \citep{PeretsBeniamini21}.

The recent discovery and detailed observations of a close by SN 2019ehk in M100 \citep{de21,nakaoka21}, shed new light on the properties of such \cas{} SNe, and in particular gave rise to the first detection of early X-ray emission from a thermonuclear SN \citep{WynnV+20b,WynnV+21}. 

Given these various properties showing evidence for (1) large He abundances (either directly observed, or  potentially inferred from the large abundance of He-burning intermediate element products), (2) inefficient burning, leading to low production of \niVI{} and faint, low luminousity SNe, and (III) old environments, we suggested that these SNe form a novel type of SNe, arising from theremeonuclear explosions of He-rich WDs \citep{Perets+10}, rather than from the core-collapse of massive stars (typically thought to be the progenitors of type Ib SNe). 

Production of faint and \cas{} SNe in models were already seen in the 1980s \citep{Woo+86}, were such SNe were produced in models of detonation of a He-shell (suggested to be accreted from a companion through stable mass-transfer) on CO WDs, which failed to explode the CO-WD. These did not receive much attention as they did not produce normal type Ia SNe, nor any other type of SNe observed at the time. More recent studies further explored these models, and termed such SNe .Ia SNe \citep{Bildsten+07}. Following our identification and characterization of \cas{} arising from He-rich WD explosions, several models tried to reproduce the properties of \cas{} SNe. These models explored the detonation of a massive He-shell on top of low-mass CO WDs \citep{She+10,Waldman+11}. Although such models did produce faint and \cas{} SNe they produced light curves and spectral series were inconsistent with the observed evolution of \cas{} SNe. Other scenarios such as a WD disrupted by a neutron star (NS) or a black hole (BH) typically produced too faint and too rapid evolving SNe, not resembling \cas{} SNe \citep{Metzger12,Fernandez&Metzger13,MacLeod+14,Zenati19b,Zenati+20a,Bobrick+21}. Classically study in the later stage of NS-He WD binaries, those may observed as ultracompact X-ray binaries (UCXBs) \citep{Nelson+86,ChenLiang+22}. Furthermore, population synthesis models of NS-WD mergers did not produce significant long delay time distributions that could be consistent with the significant fraction of early type host galaxies for \cas{} SNe \citep{Toonen+18}.

We have already proposed earlier \citep{Perets+19} that hybrid HeCO WDs \citep{ibentutukov85,Han+07,PradaStraniero09,Zenati+19a} play a key role in the production of theremonuclear SNe \citep{Perets+19,Pakmor+21}, and may produce most of the normal type Ia SNe \citep{Perets+19}. Here we propose a novel model for the origins of \cas{} SNe, in which a hybrid HeCO WD  disrupts a low-mass CO-WD. The debris dynamically accrete on the hybrid WD, and then the CO debris mixes with the He-shell of the hybrid, and detonates, while the hybrid WD core does not explode. We show that such scenario can reproduce the overall light curve and spectra of \cas{} SNe. We also find large abundances of Ca and \TiIV{} are produced. As we also discuss elsewhere, the progenitors of such disruptions give rise to a significant population of SNe with long delay times that would be observed in early type host galaxies as observed for \cas{} SNe (Toonen et al., in prep.).

{} This paper is organized as follows. In Section \ref{Sec:model} we overview the proposed scenario and its basic aspects. We then present the initial conditions and structure of the post-disruption debris disk in our models \ref{sec:Initial-conditions}.  In the methods section \ref{sec:methods} we present the details of our numerical results, including the \flash{} axisymmetric hydro-thermonuclear simulations and the radiative transfer modeling. In Section \ref{Sec:Results} we present the disk evolution and explosion seen in our models, and the resulting light curves and spectra. We then discuss (section \ref{Sec:discussion}) our results in comparisons with observations, explore their implications, and point to potential caveats. In Section \ref{Sec:PPS} briefly shows the stellar populations of \cas{} followed by our summary \ref{Sec:conclusion}.

\section{Disruption of a CO-WD by a HeCO WD}    \label{Sec:model}
Hybrid HeCO WDs \citep{ibentutukov85,Han+07,PradaStraniero09,Zenati+19a} are formed in interacting binaries where an evolved red-giant is stripped of its hydrogen (mostly) and partly stripping of helium-rich envelope during its evolution. This complex binary evolution channel can give rise to hybrid-WDs, composed of significant fractions of both CO and He. Such hybrid WDs reside in the mass range of $\sim0.4-0.73$ $\rm M_{\odot}$ and contain a He-envelope containing $\sim2-20\%$ of the WD-mass \citep{Zenati+19a}, with lower mass hybrids typically containing larger He fractions (but other HeCO formation branches might produce larger He fractions even in the high mass range; \citealt{Pakmor+21}). Such hybrid-WDs are frequent among compact WD-WD binaries that inspiral and merge in a Hubble-time; where $25\%$ of all WD-WD mergers may include a HeCO-WD \citep{Perets+19}. 

In a previous study we explored the case of a CO WDs of $>0.7$ M$_\odot$ disrupting lower mass HeCO WDs, and eventually producing normal type Ia SNe \citep{Perets+19}. Here we investigate the cases where low-mass CO-WDs are disrupted by hybrid HeCO-WDs, and show that these lead to the production of \cas{} SNe with detailed properties matching those of observed \cas{} SNe.

In the current models we do not simulate the actual disruption of the lower-mass WD, which require an expensive and less resolved 3D simulations,  but assume, similar to \citep{Perets+19} that it was already disrupted and formed a debris disk around the higher mass WD. The axisymmetric disk now allows us to use a 2D model. We therefore introduce a disk of debris around a HeCO in a 2D axis-symmetric simulation and follow the accretion evolution of the debris onto the embedded HeCO WD. As discussed below, the accreted material eventually heats up and a He-enriched detonation ensues in the outer shell of the HeCO WD, leading to a weak explosion, which leaves the core of HeCO WD intact. We then model the light-curve and spectra produced from the explosion using a non-LTE 1D code \texttt{CMFGEN}.
In the following we discuss the various assumptions and modeling processes in more detail.

\section{Initial conditions of the post-disruption debris disk}
\label{sec:Initial-conditions}

\subsection{Disk structure}
We focus on disks that form when a CO WD is tidally disrupted by a more massive HeCO-WD companion in a close binary system, and follow similar modeling of debris disk evolution around compact objects as done by us and others in other contexts  (See, \citealt{Papaloizou&Pringle84,Fryer+99,Metzger12, Fernandez&Metzger13,Bobrick+17, Zenati19b, Perets+19, Metzger+21}). In this section we estimate the  characteristic properties of these binary mergers analytically, and provide insights onto the initial conditions of the debris disk, the transient WD accretion phase, and the key timescales involved. 
We use these analytic results to establish the initial conditions for our numerical simulations described in Section \ref{sec:methods}.

\begin{table}
  \begin{center}
    \caption{Simulation Suite}
    \begin{tabular}{|c|c|c|c|c|c|c|c}  
     \hline
      Model & $M_{c}$ & $M_{d}$ & $\alpha$ & $R_{\rm d,0}$ & $t_{\rm visc,0}^{(a)}$ \\
      \hline 
      - & ($M_{\odot}$) & ($M_{\odot}$) & - & (cm) & (s) \\
\hline
\hline 
fca$_1$ & 0.63 & 0.55 & 0.1 & $2.08\times 10^{9}$ & $53.2$ \\
fca$_2$ & 0.63 & 0.52 & 0.1 & $2.2\times 10^{9}$ & $61.8$ \\
fca$_3$  & 0.58 & 0.52 & 0.1& $2.1\times 10^{9}$ & $50.1$ \\
fca$_4$ & 0.63 & 0.55 & 0.05 & $2.08\times 10^{9}$ & $106.4$ \\
\hline
\end{tabular}
\label{tab:models}\\
{\bf Notes:} $^{(a)}$Calculated from Eq.~(\ref{eq:tvisc}), assuming a disk aspect ratio $\theta = 0.8$. Model fca$_4$ was not studied with tracer particles.
\end{center}
\end{table}

\subsubsection{Disk Formation}
The disruption process of the companion WD depends on the stability of the mass loss following the onset of Roche lobe overflow (RLOF). Mass transfer occurs as the binary loses orbital angular momentum, resulting in RLOF of the secondary onto the primary. 
We are interested in the fate of unstable mass transfer in a binary system consisting of a HeCO WD primary of mass $M_{\rm WD}$ and radius $R_{\rm WD}$ orbited by a secondary CO WD companion of mass $M_{\rm WD_{CO}} \lesssim M_{\rm WD_{HeCO}}$ and radius $R_{\rm WD_{HeCO}} \gtrsim R_{\rm WD_{CO}}$. 

The characteristic radial dimension of the disk can be estimated as (e.g.,~\citealt{MargalitMetzger16})
\begin{equation}
R_{\rm d,0} = a_{\rm RLOF}(1 + q)^{-1}.  \label{eq:Rc}
\end{equation}

The mass of the formed disk will be about equal to the original CO-WD, $\rm M_{\rm d,0} \approx M_{WD}^{CO}$. 
The stability of the mass-transfer depends on the mass ratio of the binary $\rm q =\frac{M_{WD}^{CO}}{M_{WD}^{HeCO}} = 0.83-0.88$.
For circular orbits, this takes place at an orbital separation \citep{Eggleton83}
\begin{equation}
a_{\rm RLOF} \approx R_{WD}^{HeCO}\frac{0.6 q^{2/3}+ {\rm ln}(1+q^{1/3})}{0.49 q^{2/3}},  \label{eq:RLOF}
\end{equation}

These provide us with the typical masses and scales of the formed disks.
Following the disruption, the newly formed disk is very thick, as also seen in 3D models of the disruption stage. The vertical scale-height ${\rm H_0}$ and aspect ratio are typically $\theta_0 \equiv H_0/R_{\rm d,0} \sim 0.55-0.9$ \citep{Metzger12,MargalitMetzger16,Zenati19b,Metzger+21}. The gravity pressure is $P_{*} \simeq \rho c_{s}^{2} \simeq{} \Sigma/{2 \pi h}$ where $\Sigma{}$ is the initial surface density of the disk and $c_{s}$ is the sound speed.

\subsubsection{Disk stability}
The formed disk evolves through accretion onto the central object, during which the disk heats up. Magnetic torques arising from the magneto-rotational instability (MRI; \citealt{BalbusHawley98}) likely transfer angular momentum of the material outward in the accretion disk so that material inspirals into the central object (in this study the HeCO WD). Our model does not include magnetic fields, and can not self-consistently introduce viscous evolution to the disk. Instead, the viscous evolution is modelled through the use of artificial viscosity, following a Shakura-Sunyaev-like viscous evolution. In principle gravitational instabilities might produce spiral arms in the the disk and provide other angular momentum transfer processes. However, the disk is likely to be stable. The \citet{Toomre64} stability parameter, Q, is given by

\begin{equation}
Q = \frac{\Omega c_s}{\pi G \Sigma} \underset{r = R_{\rm d,0}} \approx \frac{(M_{WD}^{HeCO}+M_{WD}^{HeCO})^{2}}{M_{WD}^{HeCO}\  M_{WD}^{CO}}\frac{H_0}{a_{RLOF}} \sim \frac{1+q}{q}\theta.  \label{eq:toomreQ}
\end{equation}

For the typical disks masses and thickness we consider here,  $\theta_0 = 0.55$ and $q \gtrsim 0.83-0.88$ $Q = 1.17-1.23$, and the disk is stable.
Generally, during the following evolution of the disk the viscous heating is larger than the radiative cooling, and the disk is likely to be gravitationally stable up to the point a detonation occurs. This should nevertheless be further verified in future 3D simulations.
Here, using 2D simulations we assume only viscous evolution; and postpone studies of other scenarios to future 3D models. In the following we briefly describe the overall timescales and accretion rates in this process, and then discuss the numerical hydrodynamical modeling in the next sections.
 
Let us consider the relevant timescales which will later refer to.
$\rm t_{\rm orb}$ is orbital time-scale,
\begin{equation}
t_{orb} \simeq {42\left(\frac{R_{d,0}}{R_{WD}^{HeCO}}\right)^{3/2} \left(\frac{M_{d,0}}{{\rm M}_\odot}\right)^{-1/2}}s,  \label{eq:torb}
\end{equation}

The viscous timescale is given by 
\begin{equation}
t_{\rm visc,0} \sim \left.\frac{r^{2}}{\nu}\right|_{R_{\rm d,0}} \sim \frac{1}{\alpha}\frac{1}{\theta_0^{2}}\left(\frac{R_{\rm d,0}^{3}}{GM_{\rm WD}^{HeCO}}\right)^{1/2}s,  \label{eq:tvisc}
\end{equation}
where $\nu = \alpha c_{\rm s}H = \alpha r^{2}\Omega_{\rm K}\theta^{2}$ is the effective kinematic viscosity, $\Omega_{\rm K} \equiv (GM_{d,0}/r^{3})^{1/2}$ is the Keplerian orbital frequency, $c_{\rm s} \approx H\Omega_{\rm K}$ is the midplane sound speed, and $\alpha$ parameterizes the disk viscosity (see also sub-section \ref{SubSec:Flash}).

The viscous accretion results in a characteristic accretion rate following \citep{Shakura_Sunyaev73}, where the characteristic peak accretion rate is approximately,
\begin{equation}
\dot{M}_0 \sim \frac{M_{\rm d,0}}{t_{\rm visc,0}} \sim 0.1 \Msun ,\alpha_{=0.1}.
\end{equation}

The timescale for photons to diffuse out of the disk midplane is then,
\begin{equation}
t_{diff} \simeq \kappa_{es}\frac{\Sigma}{\sqrt{2}\pi}H_{0} \simeq{} 1-7\times 10^5 yr,  \label{eq:tdiff}
\end{equation}
and is typically longer than the other timescales, such that radiative cooling does not play an important role. 

We therefore consider a stable accretion disk around the primary HeCO WD, and follow its evolution through 2D hydrodynamical simulations as we discuss below.

\subsection{Disk structure}
The initial disk model can be calculated using  energy and pressure considerations.
The total energy in the disk is given by the combined contribution from the gravitational. thermal and kinetic energy, $ e_{\rm tot}= E_{\rm thermal}+E_{\rm kinetic+rotation}+E_{\rm gravity} $, given by
\begin{equation}
\label{eq:etot}
e_{tot} = \frac{1}{2} [v_{\bar\rho}^{2} + \frac{l_{z}^{2}}{\bar\rho^{2}}] -\frac{GM_{d,0}}{r} - E_{g}(r,\rho) + e_{int} . 
\end{equation}

The maximum density $\rho_{\rm max}$ and temperature $T_{\rm max}$ are expected at the initial innermost part of the disk, $R_0$, (in the midplane $z=0$). At that position the pressure gradient must vanish:
	\begin{equation}
		\nabla p = 0 \mskip20mu \text{at} \mskip20mu \bar{\rho} = R_0, \label{eq:grad_press}
	\end{equation}
	where $p = p(\rho,T,X_i)$ is determined either using an ideal gas plus radiation or the Helmholtz EOS. 

Our solution of the disk geometry depend on the EOS and the equilibrium for the fluid which is dominant by HeCO WD potential $\phi_c$, the centrifugal forces, and the specific angular momentum \citep{Papaloizou&Pringle84}.

\begin{equation}
		-\frac{1}{\rho} \nabla{P} -\nabla \phi_c +\Omega^2 \bar\rho \hat{{\bar\rho}} = 0. \label{eq:diskequ_profile}
\end{equation}

By combining the above equation with equation (\ref{eq:grad_press}) and the EOS, we can deduce the general equation (\ref{eq:diskequ_profile}) following \citep{Stone+99}, see also \citep{ Fernandez&Metzger13} and \citep{Zenati19b}, 

\begin{equation}
	\frac{P}{\rho} = \frac{(\gamma +1)GM_{c}}{\gamma R_0}\left[ \frac{R_0}{r}-\frac{1}{2}\left(\frac{R_0}{r\sin\theta}\right)^2 - \frac{1}{2d}\right]. \label{eq:diskpressure}
\end{equation}

We can then use the conditions \eqref{eq:etot} and \eqref{eq:grad_press}, and the temperature corresponding to $\frac{\gamma \Re_{g} T}{M_{d,0}}=\frac{GM_c}{R_0}$,  where $\Re_{g}$ is the gas constant. In addition we can solve for $T_{\rm max}$ and $\rho_{\rm max}$.
Following \citet{Stone+99} we normalize the torus density by $\rho_{\rm max}$ from \ref{eq:etot} and \ref{eq:grad_press}, resulting in the density distribution
	\begin{equation}
		\frac{\rho}{\rho_{\rm max}} = \left\{ \frac{2d}{d-1}\left[ \frac{R_0}{r}-\frac{1}{2}\left(\frac{R_0}{r\sin\theta}\right)^2 - \frac{1}{2d}\right]\right\}^{\frac{1}{\gamma -1}}. \label{eq:torus_rho_profile}
	\end{equation}
	where d is the torus distortion parameter which is related to the internal energy at $R_0$ by
	\begin{equation}
	    e_{\rm int,max} = \frac{GM_{\rm HeCO}}{R_0} \frac{1}{2\gamma}\frac{d-1}{d}. \label{eq:e_int_d}
	\end{equation}
	
	Bounded torus configurations require distortion parameter $d>1$. Imposing the requirement $d>1$ for physical solutions, one can obtain $d$ from the condition \eqref{eq:e_int_d} using $T_{\rm max}$ and $\rho_{\rm max}$ from \ref{eq:etot} and \ref{eq:grad_press}.
	Here, we set $\gamma=5/3$ as an initial value, and then use iterative computations of the effective adiabatic index using the Helmholtz EOS, until a steady configurations is achieved. 
	
	Equipped with the initial disk structure, we can now explore the evolution of the disk and its explosive outcomes.

\section{Methods}        
\label{sec:methods}
We model the evolution of the debris disk through a 2D thermonuclear-hydrodynamical modelling using the  \flash{} open code \citep{Fryxell+00}, with a 19 elements nuclear network. The initial properties of the WDs and the disk compositions are calculated from stellar evolution 1D models of HeCO WDs \citep{Zenati+19a} which are mapped into 2D in FLASH. We post-process the results from the explosions through a larger nuclear network using the TORCH code \citep{Timmes2000} in MESA \citep{Paxton+15}. We then make use of the non - local thermodynamic equilibrium (non-LTE) radiative transfer code \texttt{CMFGEN} \citep{HD12} to model the light-curve and spectra arising from the modelled explosions, which we then compare to the observations. These various steps are discussed in more detail below. 

The nature of our \flash{} simulations is similar to those we ran earlier in other WD (or neutron star/black-hole) + disk configurations, and are described in  \citep{Zenati19b,Perets+19,Zenati+20a,Zenati+20b,Bobrick+21,Metzger+21}. However, the type and masses of the WDs involved and their compositions differ from our previous models, leading to very different outcomes. The modeling details are described below.

\subsection{Numerical modelling through hydrodynamical simulations}
\label{SubSec:Flash}
We simulate the evolution of the HeCO-WD merger with another low-mass CO-WD, by taking the latter to be an accretion torus around the former. As discussed above, the debris disks from the CO WDs were modelled following the same procedure used by us earlier \citep{Perets+19}, and discussed above, where we assume the disk composition is fully mixed, and follows the composition of the disrupted CO-WD as found in our stellar-evolution models. The HeCO WDs profiles were taken from our 1D stellar evolution modeling using \texttt{MESA} stellar evolution code to model the density, temperature and compositions profiles of HeCO and CO WDs as found in our previous study \citep{Zenati+19a}, which were then mapped to the 2D configuration in FLASH. 

 The initial conditions (see Table \ref{tab:models} for the properties of the different runs) were used to run our detailed hydrodynamical simulations, using the publicly available \flash{} v4.3 code \citep{Fryxell+00}. In our models we employ the unsplit ${\rm PPM}$ solver of \flash{}, necessary for handling artificially generated shocks, which could spread out over a few zones by the $\rm PPM$ hydrodynamics solver, and otherwise might lead to unphysical burning within shocks. Our simulation are in ${\rm 2D}$ axisymmetric cylindrical coordinates $[{\bar\rho},z]$ on a grid of size ${\rm 2\times 1\left[10^{10}cm\right]}$, using an adaptive mesh refinement with a maximum of 12 levels (the finest) and a smallest grid cell of 15\,km. We used a reflective boundary conditions on the symmetry axis.

We follow similar approaches as described in other works on thermonuclear SNe (e.g. \citealt{Meakin+09,Perets+19}) to follow the nuclear burning and detonation throughout the evolution.

 We explore four different combinations of HeCO WD - CO WD mergers, in which the primary HeCO-WD disrupts the secondary; a lower mass CO-WD, forming a debris disk.  Although the debris disk from the disrupted CO-WD can be initially clumpy and/or asymmetric, it rapidly evolves into a relatively symmetric accretion disk around the more massive WD before any significant nuclear burning occurs \citep{Dan+12,Kashya+15}. We therefore model the mergers starting only following the formation of a symmetric debris disk around the more massive HeCO-WD in our simulations, similar to the approach used by us and others \citep{Dan+15,Fernandez&Metzger13,Zenati19b,Zenati+20b} to model mergers of WDs with WDs/NS. Such cylindrical symmetry of the disk and the central accreting WD allows us to model the merger in 2D. Such models are highly advantageous as they allow for high resolution simulations, with relatively little numerical expense in comparison with 3D simulation, but they still capture most of the important multi-dimensional aspects of the merger.

Each of our hydrodynamical simulations, includes the self-gravity of the disk, and accounts for the centrifugal force as a source term. We also include neutrino heating, though it is unlikely to play an important role for the temperatures and densities reached in our models. We solve the Euler equations in axisymmetric cylindrical coordinates $[{\bar\rho},z]$. In order not to encounter numerical issues with empty cells in the simulations, we introduce a minimal density and temperature background, which are typically taken to be $\rho_{\rm bg}= 10^{-6} gcm^{-3}$, and $\rm T_{bg} = 10^{3}\ K$, and in any case, a density which is at least 100 times smaller than the density of the debris disk, and temperature which is at least 1000 times smaller.

In order to prevent the production of artificial unrealistic early detonation that may arise from insufficient numerical resolution, we applied a burning-limiter approach following \citep{Kushnir+13}.
We use a detailed EOS and account for self-gravity, this is included as a potential multipole expansion of up to multipole $l_{\rm max}=12-32$ using the new \texttt{FLASH} multipole solver of the disk. 

We use the super time-steps method in \texttt{FLASH} 4.0 for calculating the adaptive time-steps, which is set according to the speed-of-light Courant–Friedrichs–Lewy (CFL) condition with a CFL factor 0.2. We adopt the optimal strongly stability-preserving third-order Runge–Kutta scheme. 

Using FLASH we solve the equations for mass, momentum, energy, and chemical species conservation,

\begin{eqnarray}
\frac{\partial\rho}{\partial t}+\nabla\cdot\left(\rho\mathbf{v}\right)&=&0,  \\
\frac{d\mathbf v}{dt}&=&\mathbf{f}_{c}-\frac{1}{\rho}\nabla p + \nabla\phi,  \\
\rho \frac{dl_z}{dt} &=& \bar\rho (\nabla\cdot\mathbf{T})_\phi \\
 \rho\frac{d e_{\rm int}}{d t}+p\nabla\cdot\mathbf v &=& \frac{1}{\rho \nu}\mathbf{T}:\mathbf{T}+\rho(\dot{Q}_{\rm nuc} - \dot{Q}_\nu), \\
 \frac{\partial \mathbf{X}}{\partial t} &=& \dot{\mathbf{X}}, \label{eq:chemical_species_evolution} \\
 \nabla^{2}\phi&=& 4 \pi G \rho + \nabla^{2}\phi_{\rm c}, \\
\mathbf{f}_{c}&=&\frac{l_{z}^{2}}{\bar\rho^{3}}\hat{{\bar\rho}}.
\end{eqnarray}

These include source terms for gravity, shear viscosity, nuclear reactions, and neutrino cooling. Here, $\mathbf{f}_{c}$ is an implicit centrifugal source term, where $l_z$ is the z-component of the specific angular momentum. Variables have their standard meaning: $\rho$, $\mathbf{v}$, $p$, $e_{\rm int}$, $\nu$, $\mathbf{T}$, $\phi$, and $\mathbf{X}=\{X_i\}$ denote, respectively, fluid density, poloidal velocity, total pressure, specific internal energy, fluid viscosity, viscous stress tensor for azimuthal shear, gravitational potential, and mass fractions of the isotopes $X_i$, with $\sum_i X_i = 1$. Furthermore, $d/dt = \partial/\partial t + \mathbf{v}\cdot\nabla$, $\phi_{\rm c}$ denotes the HeCO-WD potential, and $
\dot{Q}_{\rm nuc}$ and $\dot{Q}_{\rm \nu}$ represent the specific nuclear heating rate due to nuclear reactions and the specific cooling rate due to neutrino emission.

We employ the Helmholtz equation of state (EOS) free energy $F(\rho, \rm T)$ in \texttt{FLASH} \citep{Timmes&Swesty00,Timmes&Arnett99}. The Helmholtz EOS includes contributions from partially degenerate electrons and positrons, radiation, non-degenerate ions, and corrections for Coulomb effects ($P_{tot} = P_{ele} + P_{pos} + P_{ion} + P_{rad} + P_{coul}$). The most important aspect of the Helmholtz EOS is its ability to handle thermodynamic states where radiation dominates, and under conditions of very high pressure. The contributions of both nuclear reaction and neutrino cooling \citep{Chevalier89,Houck&Chevalier91} are included in the the internal energy evolution calculations, and the Navier-Stocks equations are solved with source terms due to gravity, shear viscosity and nuclear reactions.

As discussed earlier, since our calculations are axisymmetric and do not include magnetic fields, we cannot self-consistently account for angular momentum transport due to the magneto-rotational instability or non-axisymmetric instabilities (e.g. associated with self-gravity). Instead, we make the common approximation of modeling the viscosity using the $\alpha$-viscosity parameterization of \citet{Shakura_Sunyaev73}, for which the kinematic viscosity is taken to be
\begin{equation}
    \nu_{\alpha}=\alpha c_{s}^{2}/\Omega_{\rm K},  \label{eq:nualpha}
\end{equation}
where $\Omega_{\rm K} = (GM_{enc}/r^{3})^{1/2}$ is the Keplerian frequency given the enclosed mass $M_{\rm enc}$ and $c_{s}$ is the sound speed.  In our models we take either $\alpha = 0.1$ or $\alpha = 0.05$ (see table \ref{tab:models}) for the dimensionless viscosity coefficient.

Nuclear burning and nucleosynthesis are followed through a $19$-isotopes reaction network \citep{Chevalier89}. 
It is included in the simulations following a similar approach to that employed by us and others (e.g. \citealt{Meakin+09,Zenati+19a,Zenati+20a,Zenati+20b}). The nuclear network used is the \texttt{FLASH} $\alpha$-chain network of 19 isotopes, which provides the source terms $\dot{\mathbf{X}}$ in Eq.~\eqref{eq:chemical_species_evolution} and adequately captures the energy generated
during nuclear burning $\dot{Q}_{\rm nuc}$ \citep{Timmes&Swesty00}. 

We made multiple simulations with increased resolution until convergence was reached. We found a resolution of $\rm 16-20 km$ to be sufficient for convergence of up to 12\% in energy.

%%%%%%%%%%%%% tracer par.
\subsection{Tracer particles and post-processing nucleosynthesis modeling}    \label{subsec:PPN_RT}
We followed the nucleosynthesis processes throughout the simulation and their key role in the evolution and the nuclear burning and explosive evolution. We made use of the $19-isotopes$ reaction network in the \flash{} hydrodynamical simulation, in order to follow the dynamics and the energetics of the explosive evolution. We then  made use of a larger network in post-processing to follow the detailed composition of the ejecta, and allow for detailed radiative transfer modeling. 

In order to do the post-processing analysis, we introduced  $\rm 8000-20000$ passive tracer particles in the various models, to track the composition, density / entropy, and temperature of the material throughout the grid, and follow the changes in these properties, where we follow the positions and velocities of the tracer particles. The tracers were initially evenly spaced every $2\times10^{8}cm$ throughout the WD and the debris disk.

Following the 2D \flash{} runs we made use of the detailed histories of the tracer particles density and temperature to be post-processed with \texttt{MESA} one zone burner \citep{Paxton+11,Paxton+15}. We employ a $\rm 125$-isotope network that includes neutrons, and composite reactions from JINA’s REACLIB \citep{RCyburt+10}. 
Similar to our previous simulations, we find that the results from the larger network employed in the post-processing analysis showed less efficient nuclear burning, and gave rise to somewhat higher yields of intermediate elements on the expense of lower yields of iron elements, similar to the results seen in our previous models and other works \citep{Dan+12,Perets+19}. 

The tracer particles and the large network post-processing were then used to generate the detailed density/composition profile of the ejecta at the end of the simulation. In turn, these data were used to produce detailed light curves and spectra of the explosions using the non-LTE CMFGEN radiative transfer code  \citep[][see more details below]{HD12}.

\subsection{Radiative transfer modeling}     \label{subsec:NLTE}
The simulations with \texttt{CMFGEN} are 1D and are based on angular-averages of the 2D simulations performed with \texttt{FLASH}. When remapping these ejecta into \texttt{CMFGEN}, we smooth the density and elemental distribution to reduce sharp gradients. The density is smoothed by convolution with a gaussian whose standard deviation is 200\,km\,s$^{-1}$. For the abundances, we smooth by running a boxcar of width 0.01\,\msun\ four times through the grid. The outer ejecta are also extended by a low density outer region with a steep density falloff in order to have an optically thin outer boundary for the radiation. The initial ejecta correspond to an epoch soon after hydrodynamical effects are over, so that the ejecta expand ballistically. At this time of a few 100 seconds, we evolve the ejecta assuming no diffusion, heating by radioactive decay, and cooling from expansion until a time of two days after explosion. At that time of two days, we remap the ejecta into \texttt{CMFGEN}. The radial grid typically employs 80 points.   

In all simulations, radioactive decay is treated for three two-step decay chains associated with \niVI{}, $^{52}$Fe, and $^{48}$Cr (six additional decay chains were considered in \citep{D15_hedet} but the corresponding isotopes are not available in the present ejecta). Decay power is treated as per normal (see, for example, \citealt{d12_snibc}) and proper allowance is made in the nonLTE equations for nonthermal ionization and excitation, as well as the additional heating term in the energy equation. For simplicity, we compute the non-local energy deposition by solving the radiative transfer equation with a grey absorption-only opacity to $\gamma$-rays set to 0.06\,$Y_{\rm e}$\,cm$^2$\,g$^{-1}$, where $Y_{\rm e}$ is the electron fraction.

The model atom used in this work includes the following atoms and ions
He\one\ (40,51), He\two\ (13,20), C\one\ (14,26), C\two\ (14,26), O\one\ (30,77), O\two\ (30,111), Ne\one\ (70,139), Ne\two\ (22,91), Na\one\ (22,71), Mg\two\ (22,65), Si\one\ (100,187), Si\two\ (31,59), Si\three\ (33,61), S\one\ (106,322), S\two\ (56,324), S\three\ (48,98), Ar\one\ (56,110), Ar\two\ (134,415), Ca\one\ (76,98), Ca\two\ (21,77), Ti\two\ (37,152), Ti\three\ (33,206), Cr\two\ (28,196), Cr\three\ (30,145), Cr\four\ (29,234), Mn\two\  (25,97), Mn\three\ (30,175), Fe\one\ (44,136), Fe\two\ (275,827), Fe\three\ (83,698), Fe\four\ (51,294), Fe\five\ (47,191), Co\two\ (44,162), Co\three\ (33,220), Co\four\ (37,314), Co\five\ (32,387), Ni\two\ (27,177), Ni\three\ (20,107), Ni\four\ (36,200), and Ni\five\ (46,183). In this list, the parentheses refer to number of super-levels and the number of full levels (see, \citealt{hm98} and \citealt{HD12} for details).

\begin{table*}
  \begin{center}
    \caption{\fca{} \flash{} simulation results}
    \begin{tabular}{c|c|c|c|c|c|c|c|c|c} 
    $\rm Model$ & $\rm M^{tot}_{ej}$ & $\rm R_{inn}$ & $\rm M_{bound}$ & $\rm M_{IME_{B}}$ & $\rm M_{IGE_{B}}$ & $\rm M_{unbound}$& $\rm M_{IME_{UB}}$ & $\rm M_{IGE_{UB}}$ & $\rm M_{E_{kin}}$ \\
    \hline 

- & $[M_{\odot}]$ & $[{\rm cm}]$ & $[M_{\odot}]$ & $10^{-4}[M_{\odot}]$ & $10^{-4}[M_{\odot}]$ & $[M_{\odot}]$ & $[M_{\odot}]$ & $[M_{\odot}]$ & $\rm E_{50}[erg]$ \tabularnewline

\hline
\hline

fca$_1$ & 0.457 & $1.5\times10^{8}$ & 0.081 & 41.24 & 0.858 & 0.377 & 0.167  & 0.022 & 1.693 \\
fca$_2$ & 0.431 & $1.7\times10^{8}$ & 0.068 & 30.17 & 0.751 & 0.364 & 0.173  & 0.052 & 1.318 \\
fca$_3$  & 0.365 & $5.2\times10^{8}$ & 0.075 & 32.16 & 1.127 & 0.288 & 0.159  & 0.012 & 0.951 \\
fca$_4$ & 0.461 & $1.4\times10^{8}$ & 0.088 & 40.03 & 0.951 & 0.373 & 0.171  & 0.023 & 1.664 \\
    \end{tabular}
 \label{tab:results}
  \end{center}
    {\bf Notes:} The total ejecta mass calculated after five viscous timescales (Eq.~\ref{eq:tvisc}). We show the bound/unbound ejecta calculated as positive Bernoulli parameter $\mathcal{B}e>0$. The total unbound mass is $\rm M_{unbound}$. The radius $R_{inn}$ designate the position location where the weak detonation occurred in the torus mid-plane. $\rm M_{IGE_{B}}$, $\rm M_{IGE_{UB}}$ are the bound and unbound iron group elements where the atomic number $Z\geq 23$, respectively. $\rm M_{IME_{B}}$, $\rm M_{IME_{UB}}$ are the bound and unbound Intermediate elements where the atomic number $Z\geq 10$, respectively. See the 125 isotopes appendix A.
\end{table*}

\begin{table*}
\caption{Properties of ejecta composition for the models for which light curve and spectra were modeled using \texttt{CMFGEN}.
\label{tab_composition}
}
\begin{center}
\begin{tabular}{
l|cc@{\hspace{1mm}}
c@{\hspace{1mm}}c@{\hspace{1mm}}c@{\hspace{1mm}}
c@{\hspace{1mm}}c@{\hspace{1mm}}c@{\hspace{1mm}}
c@{\hspace{1mm}}c@{\hspace{1mm}}c@{\hspace{1mm}}
}
\hline
         Model  &         C &         O &        Ne &        Si &        Ca &        Ti &      $^{48}$Cr & $^{52}$Fe & \niVI{}  \\ 
         & [\msun]  & [\msun]    & [\msun]   & [\msun]   & [\msun]   & [\msun]  & [\msun]   & [\msun]   & [\msun]     \\
\hline
fca$_1$  &  1.01(-1) &  9.91(-2) &  1.24(-2) &  2.32(-3) &  1.40(-1) &  7.97(-3) &  1.01(-5) &     0     &  2.12(-2)  \\ 
fca$_2$  &  9.38(-2) &  9.09(-2) &  1.17(-2) &  3.34(-2) &  1.09(-1) &  1.48(-2) &  1.97(-6) &  6.86(-5) &  5.00(-2)  \\ 
fca$_3$  &  8.66(-2) &  1.07(-1) &  5.00(-4) &  8.93(-4) &  1.36(-1) &  7.35(-3) &  3.24(-4) &  5.42(-4) &  1.05(-2)  \\ 

\hline
\end{tabular}
\end{center}
{\bf Notes:} Number in parentheses denote powers of ten.
\end{table*}

\begin{figure*}
\includegraphics[width=0.49\linewidth]{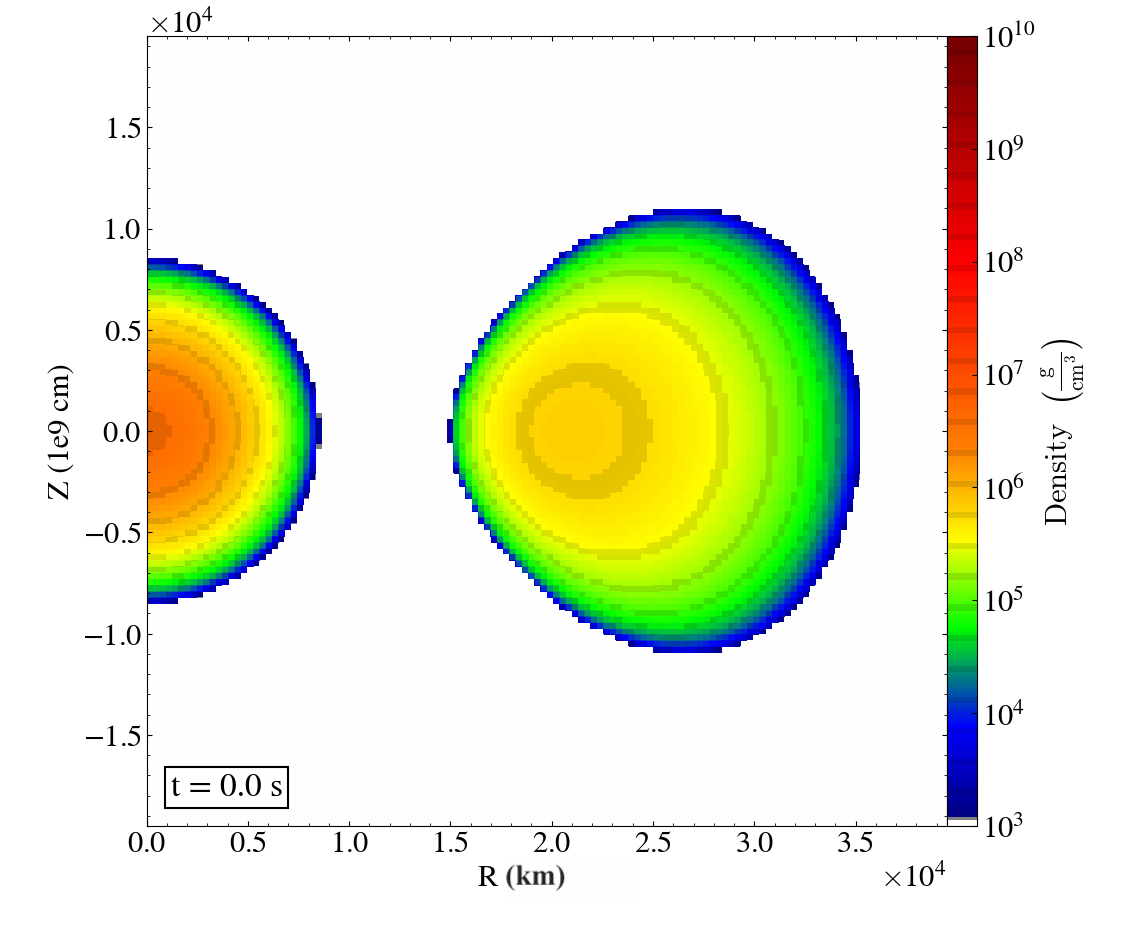}
\includegraphics[width=0.49\linewidth]{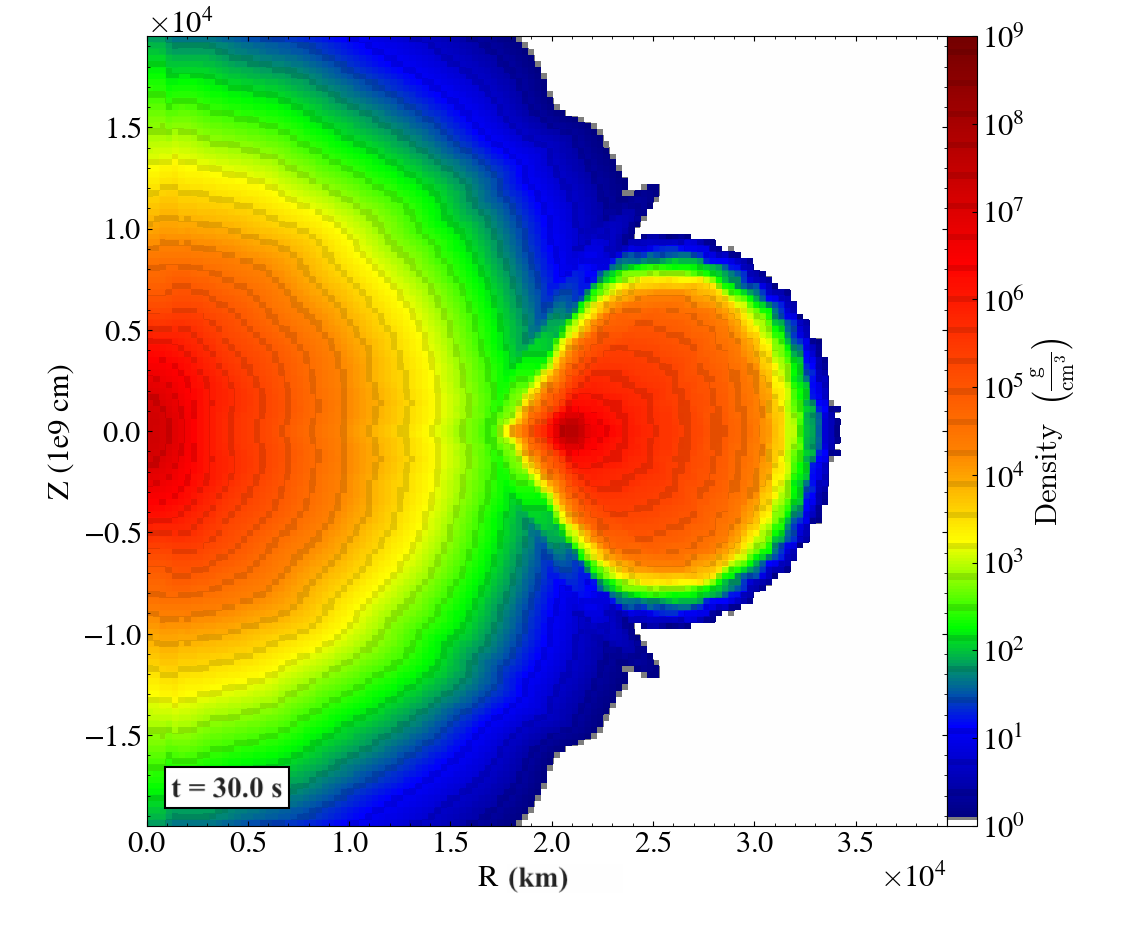}
\includegraphics[width=0.49\linewidth]{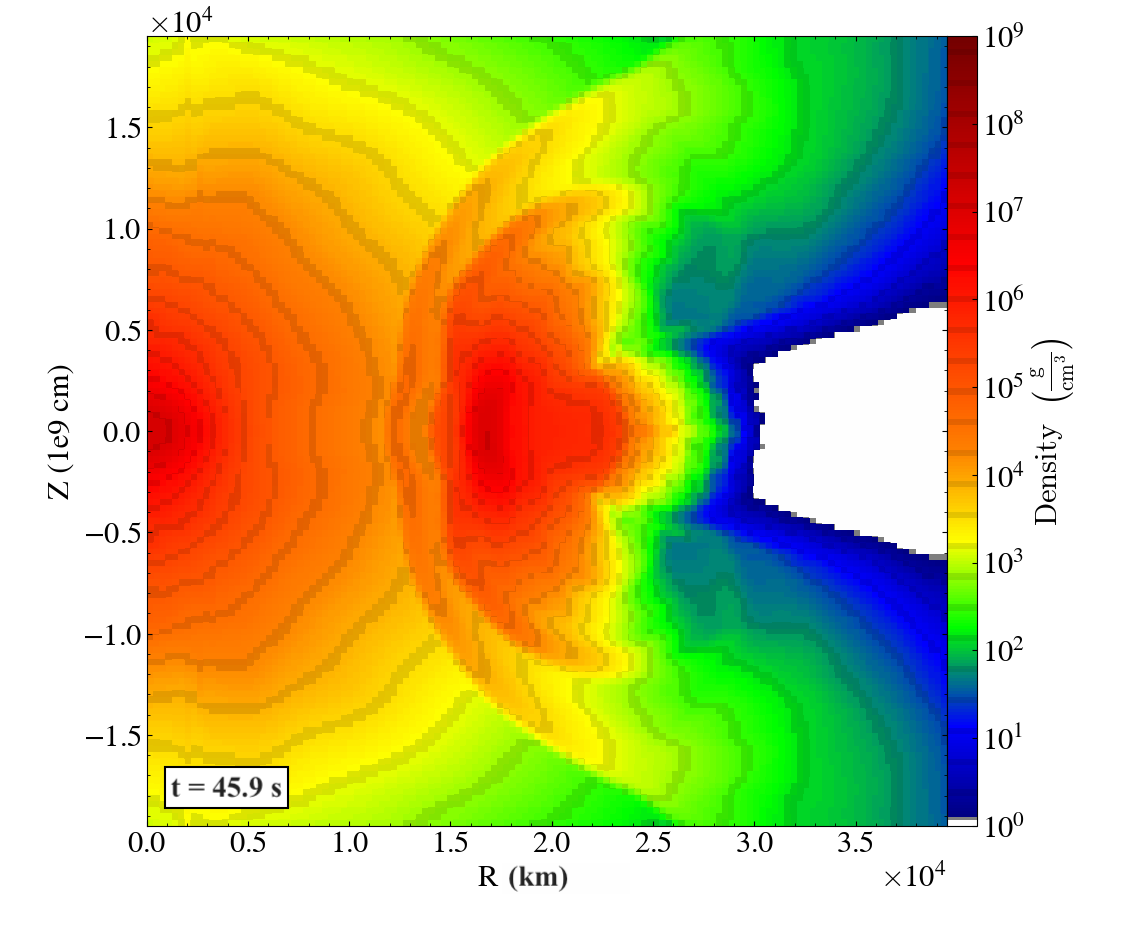}
\includegraphics[width=0.49\linewidth]{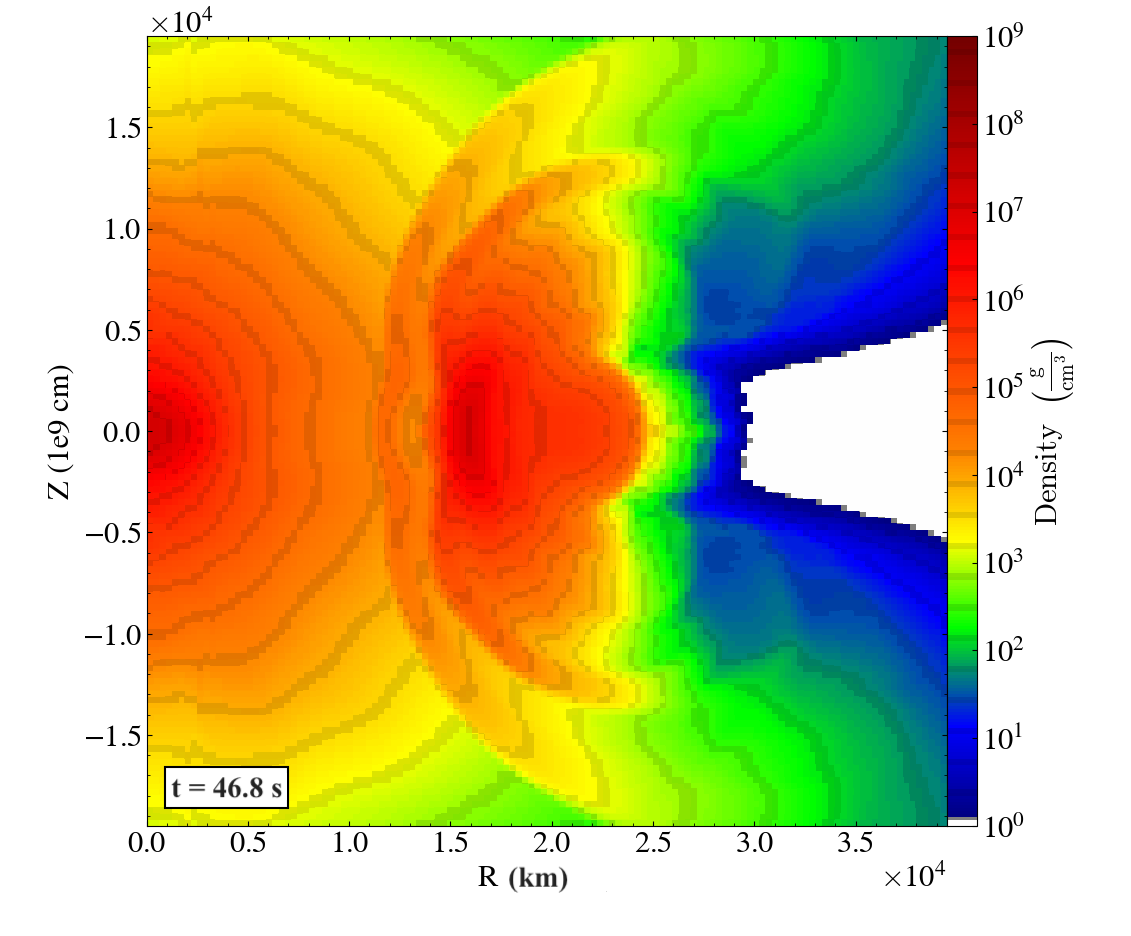}

\caption{The evolution of the debris disk from the $0.55$ $M_{\odot}$ CO-WD debris disk around the HeCO-WD of mass $M= 0.63 M_\odot$  (see table \ref{tab:models}; model fca$_1$). The panels show the color coded density distribution throughout the simulation. We show the initial model followed by several snapshots. 
Significant nuclear burning begins already at $\sim$30 s (see also figure \ref{fig:flashSi28_Ni56}), and The detonation occurs at $t_{det} \simeq{46\ sec}$, which is about 88\% of $\rm t_{visc}^{0}$. }
\label{fig:FLASH}
\end{figure*}

\begin{figure*}
\includegraphics[width=0.98\linewidth]{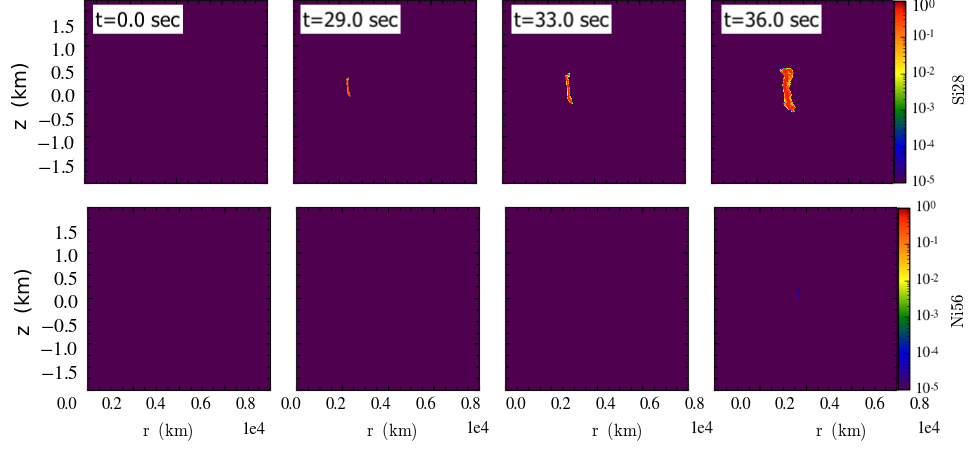}
\includegraphics[width=0.98\linewidth]{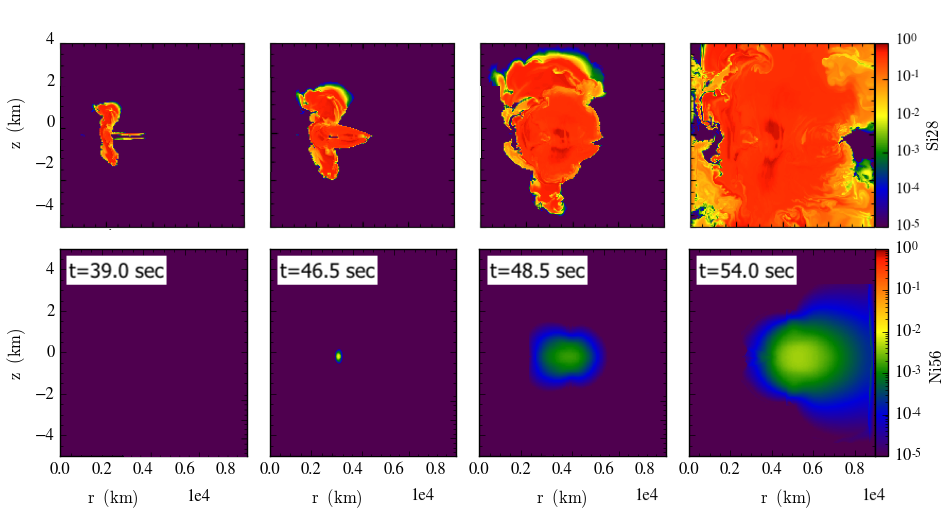}
\caption{The evolution of model fca$_1$; see table \ref{tab:models}. The two panels show the $^{28}Si$ and $^{56}Ni$ densities as a function of time. $^{28}Si$ production can be first seen after $\rm t=0.55\ t_{visc}^{0}$ followed by $^{56}Ni$ production 0.4 sec after the detonation occurred. The Z-axis (upper panel) show the zoom in fluid material.}
\label{fig:flashSi28_Ni56}
\end{figure*}

\begin{figure*}
\includegraphics[width=0.99\linewidth]{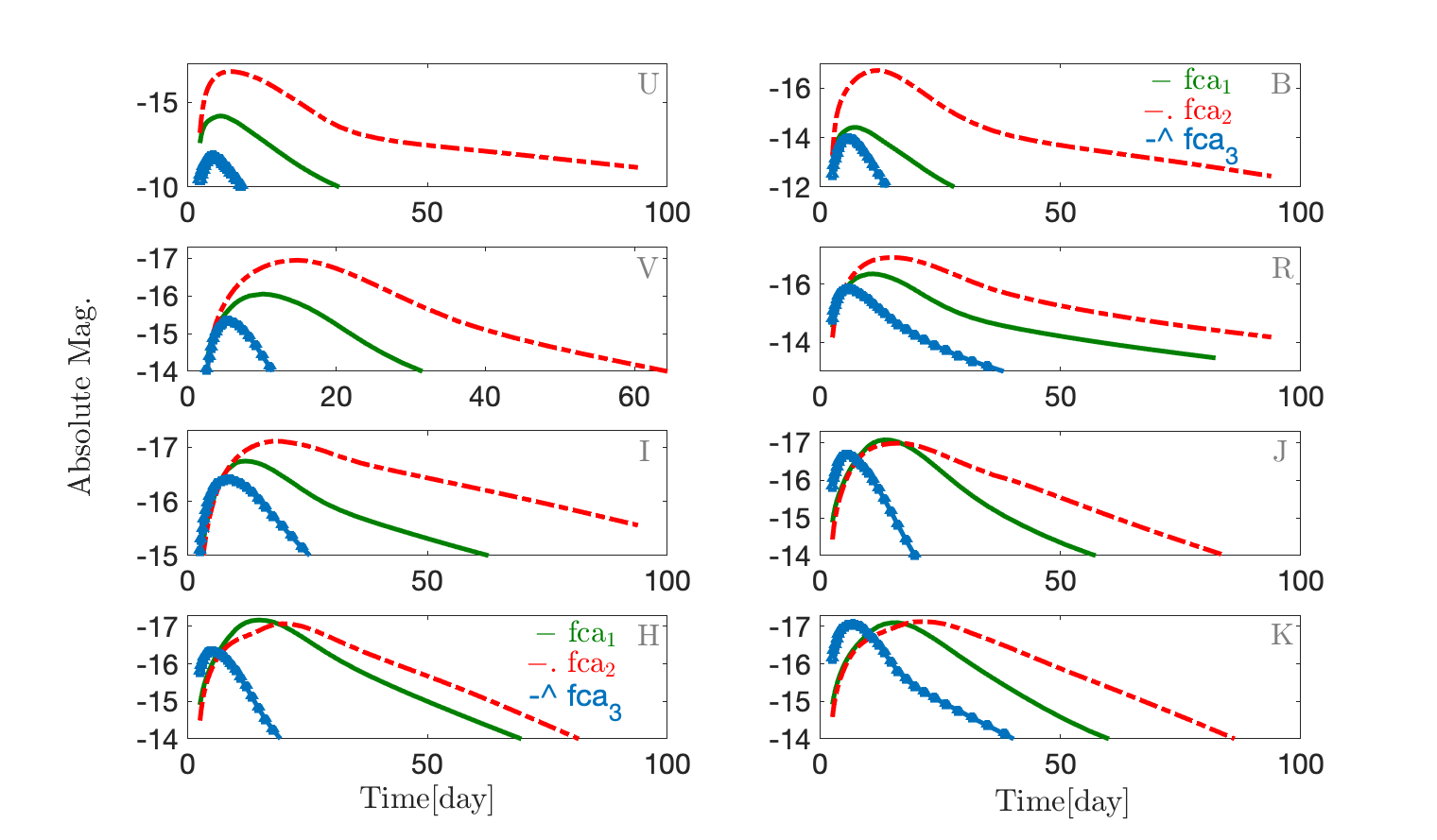}
\caption{The light curves at different  bands for the first three models in table \ref{tab:models}.}
\label{fig:Allbands}
\end{figure*}

\begin{figure*}
\includegraphics[width=0.48\linewidth]{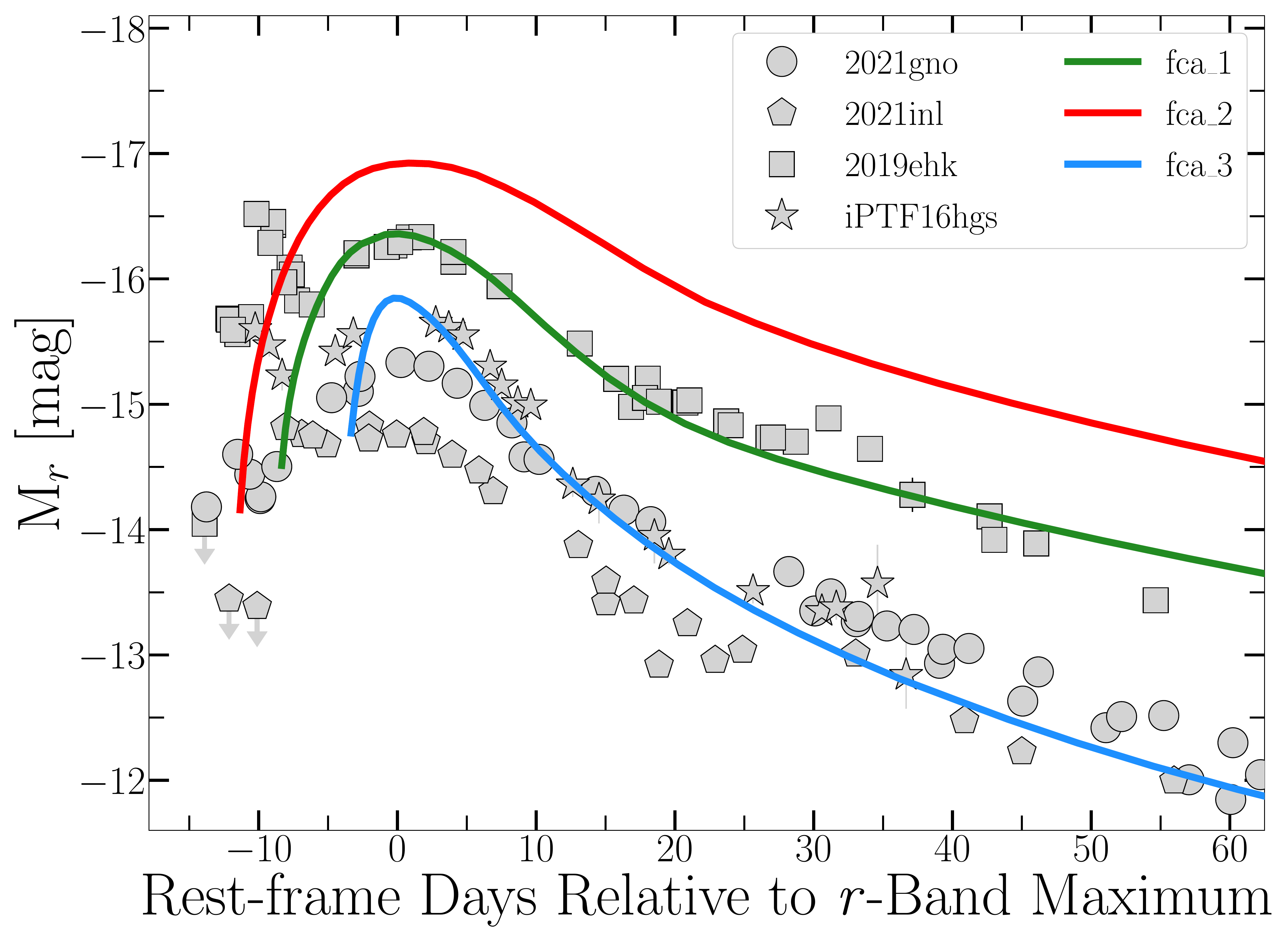}
\includegraphics[width=0.48\linewidth]{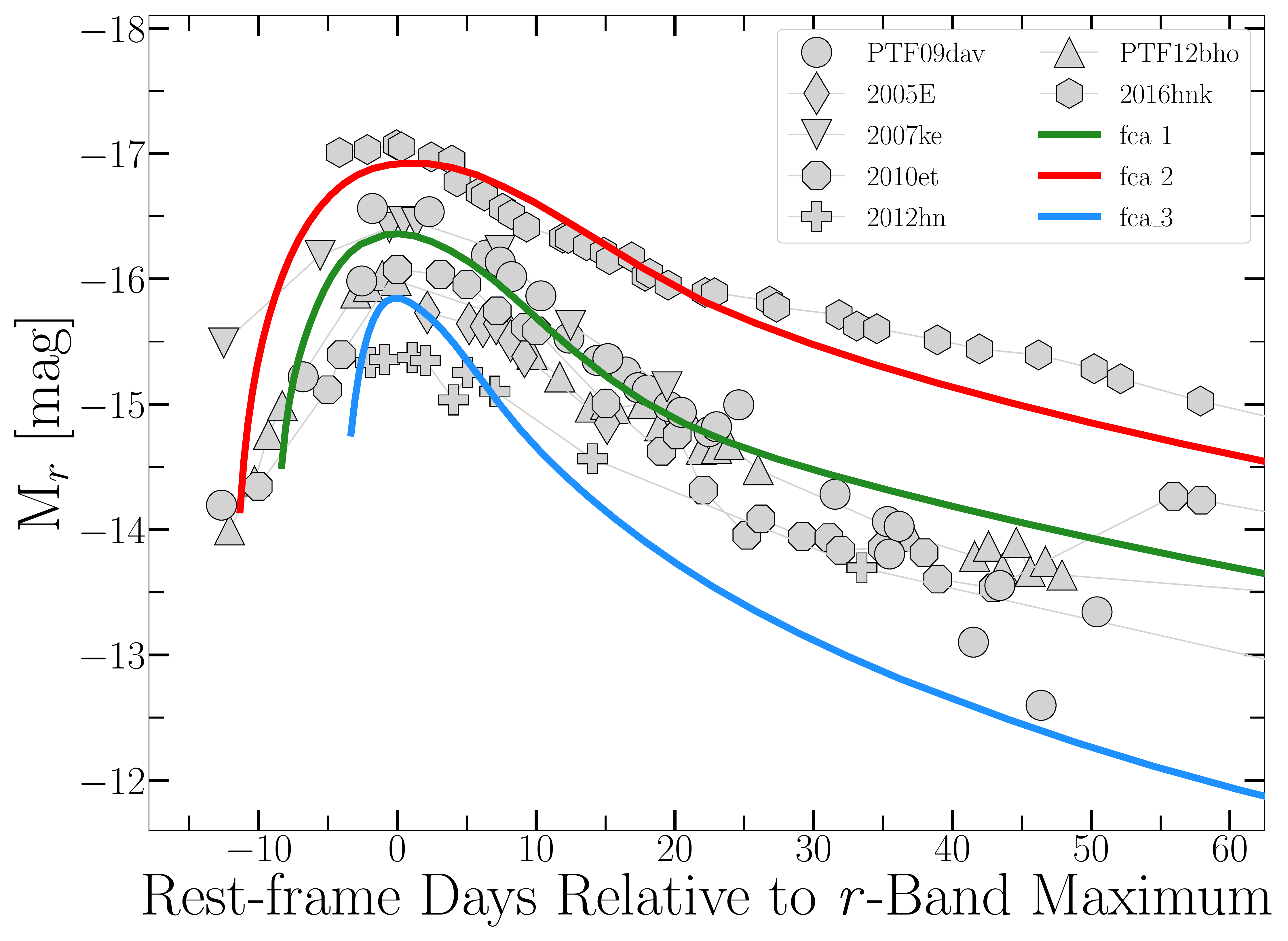}
\caption{Left/Right: Early-time absolute r-band light curves in the AB magnitude system of fca$_1$ (green curve), fca$_2$ (blue curve) and fca$_3$  (red curve) with respect to other classified \cas{} transients \citep{Perets+10, sullivan11, Kasliwal+12, lunnan17, De+18, WynnV+20b, WJG22}. Peculiar, “calcium-strong” SN2016hnk \citep{galbany19,wjg19} is also presented for reference (polygons). The \fca{} fca$_3$  model lies between the double-peaked light curves of SN2019ehk (squares), iPTF16hgs (stars), and 2021gno (circle), as well as the light curve of PTF12bho (plus). Overall, the fca$_2$ model is most consistent with the light curve evolution of SN2016hnk and the fca$_1$ model best matches SN~2019ehk's light curve.}
\label{fig:Rbandlc}
\end{figure*}

\begin{figure*}
\includegraphics[width=0.48\linewidth]{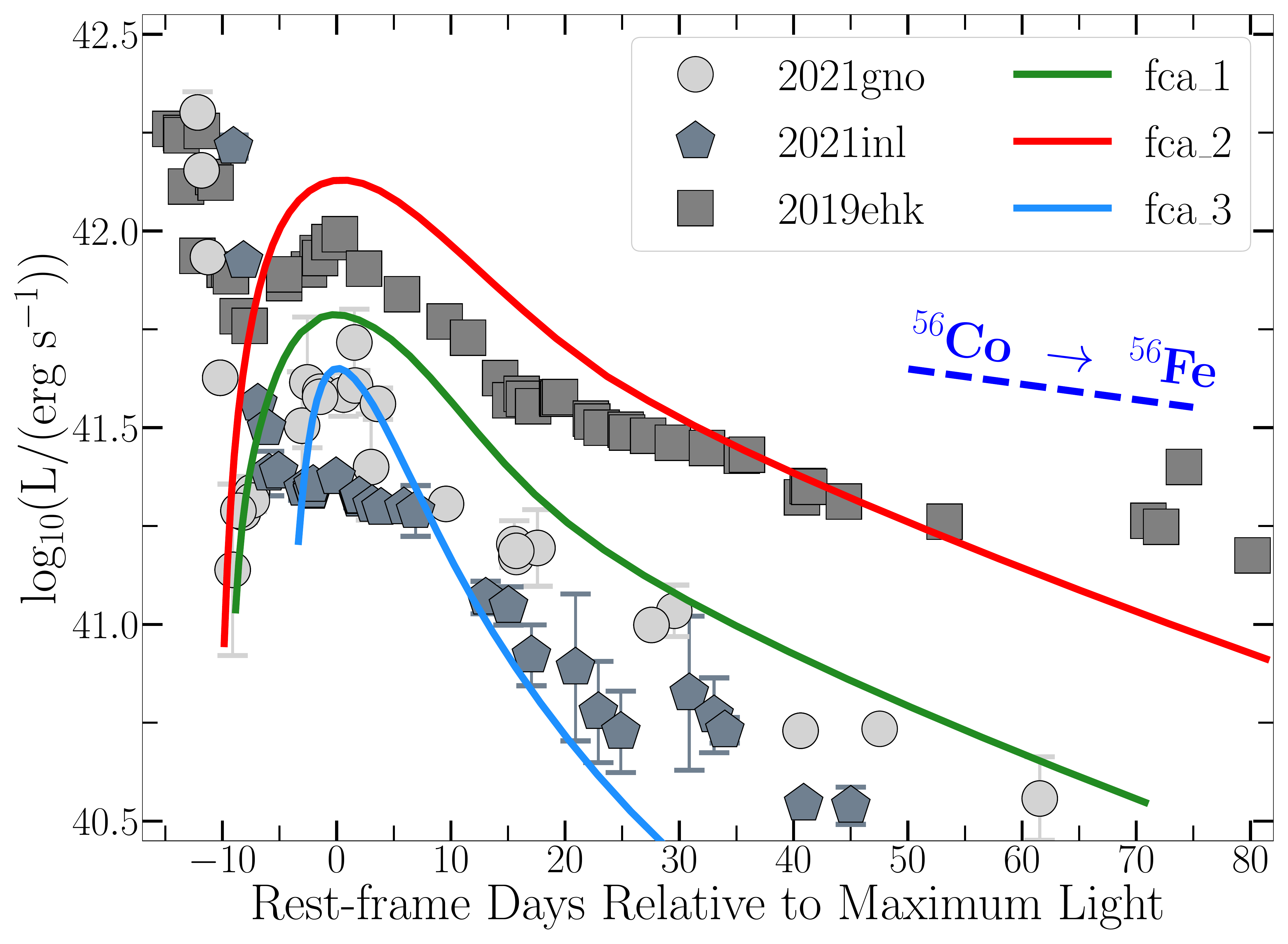}
\includegraphics[width=0.45\linewidth]{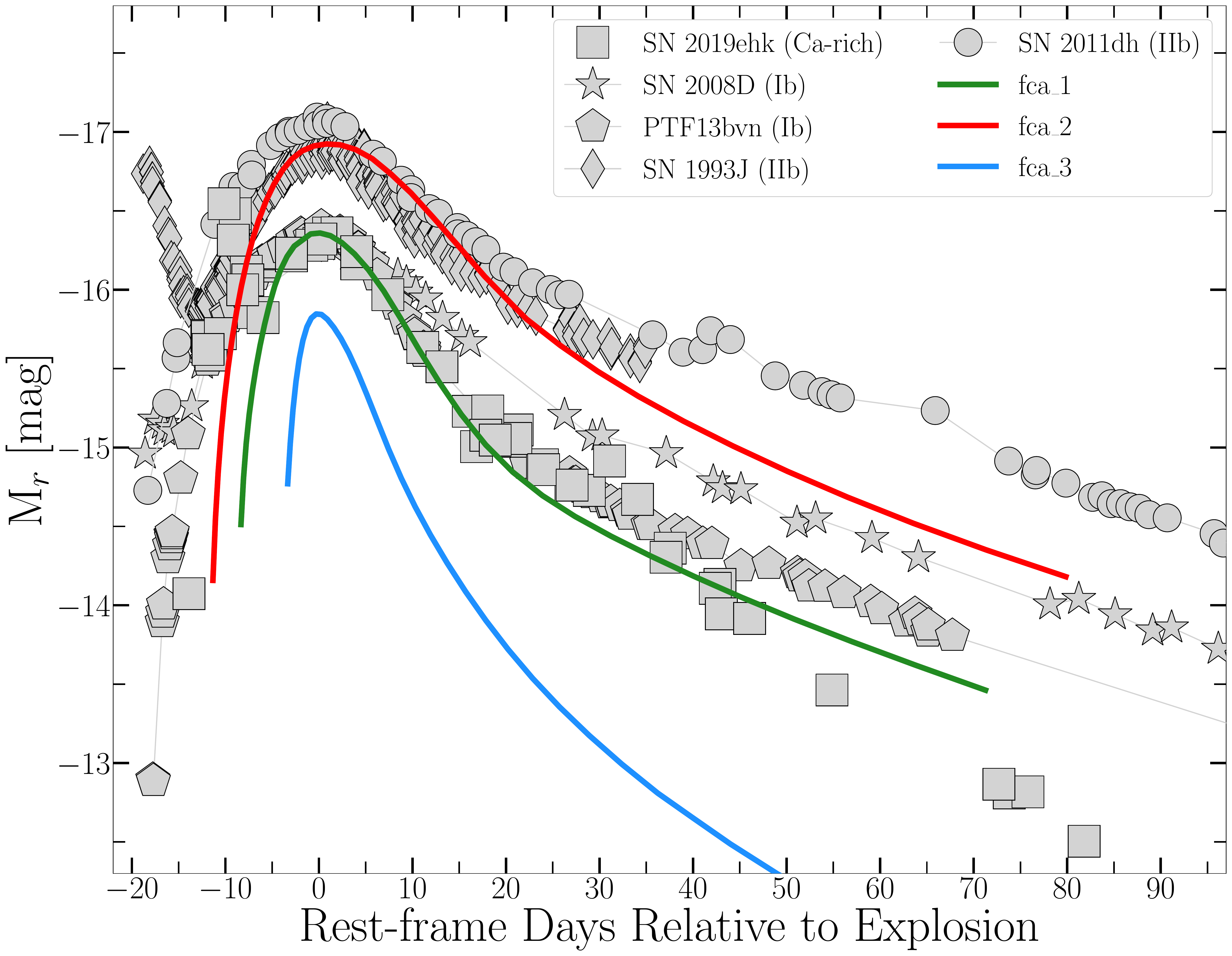}

\caption{Left panel: The bolometric light curve of fca$_1$ (green curve), fca$_2$ (blue curve), fca$_3$  (red curve) compared to \cas{} SNe ~2019ehk (squares), 2021gno (circles) and 2021inl (polygons). Right panel: comparison of the three models in the r-band $\rm M_{r}$ to type Ib/IIb SNe and SN~2019ehk relative to maximum light in the AB magnitude system.}
\label{fig:lcBol_IIb}
\end{figure*}

\begin{figure*}
\includegraphics[width=0.48\linewidth]{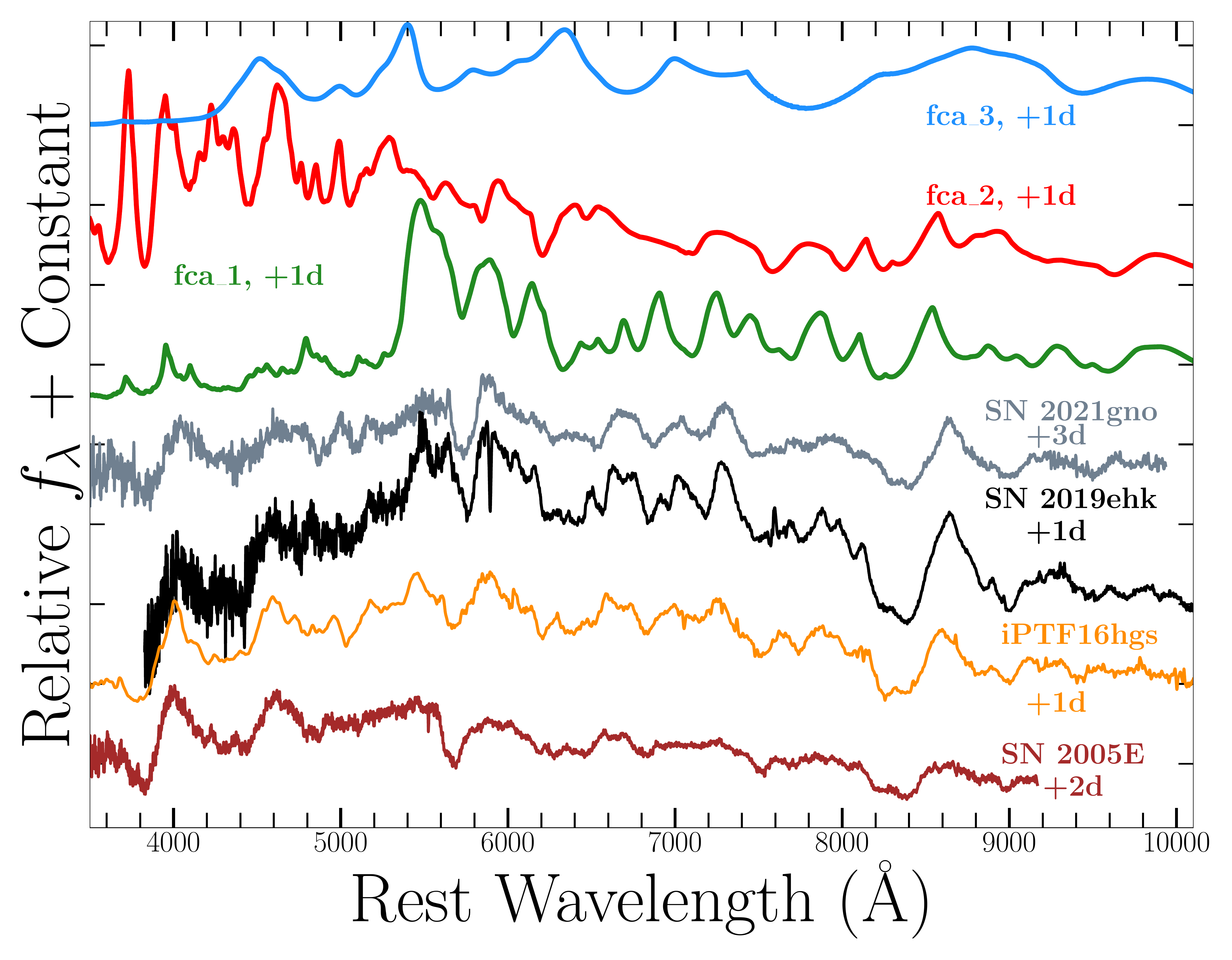}
\includegraphics[width=0.48\linewidth]{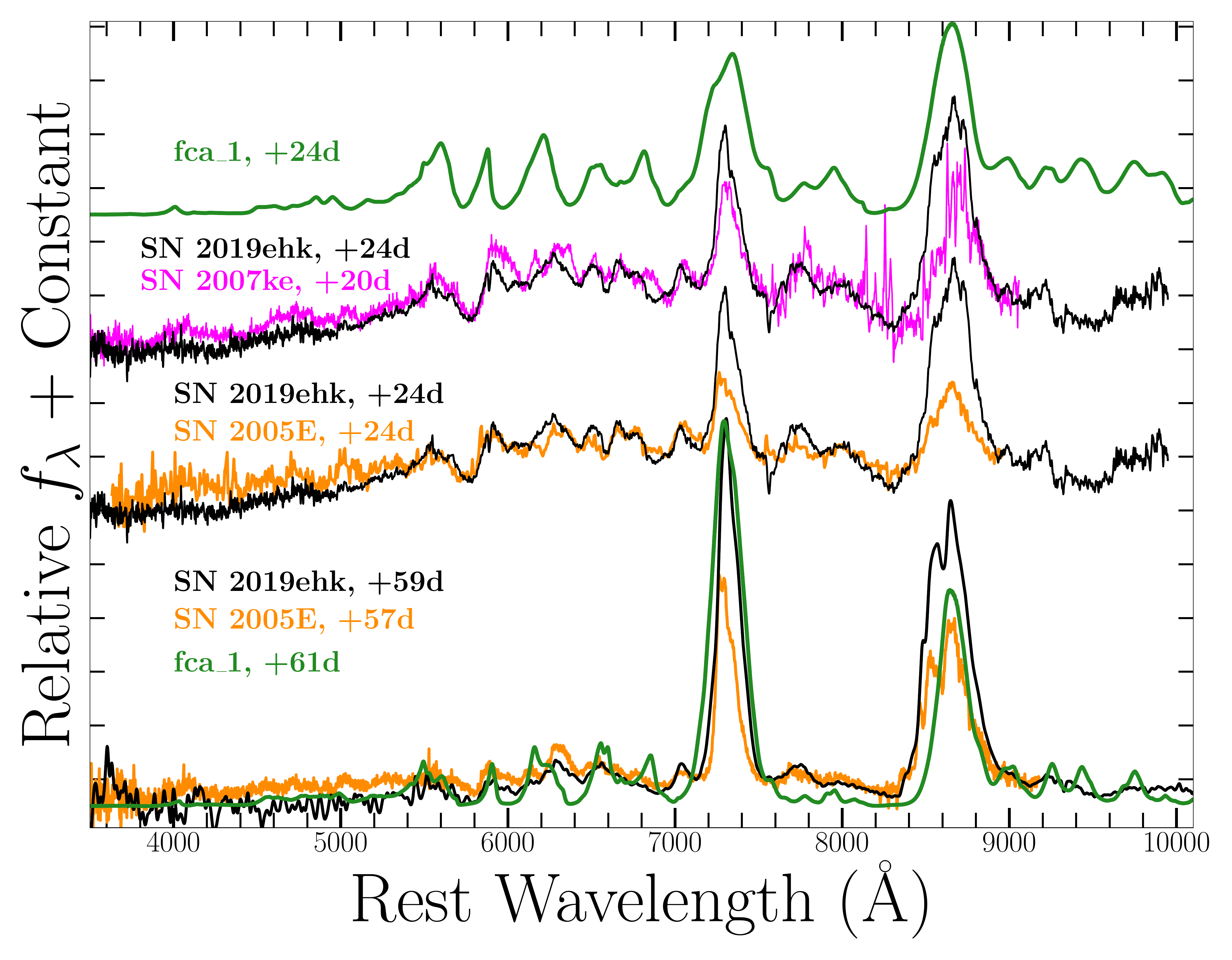}

\caption{Left panel: Early-time spectral comparison of models fca$_1$ (green), fca$_2$ (red) and fca$_3$  (blue) with the observations of SN2019ehk (black) and other \cas{} SNe at approximately the same phase. Model spectra are normalized to match the continuum of observed \cas{} object spectra. Right panel: Direct spectral comparison of fca$_1$ to SN2019ehk (black) and \cast{} SN 2007ke (magenta) and SN2005E (orange) at approximately the same phase. Model spectra are normalized to match the Ca II emission line profiles observed in SN~2019ehk. Almost every line transition is matched between spectra, with our model fca$_1$ showing stronger [Ca II] emission than SN~2005E.}
\label{fig:spectraNLTE}
\end{figure*}

\begin{figure*}
\includegraphics[width=0.96\linewidth]{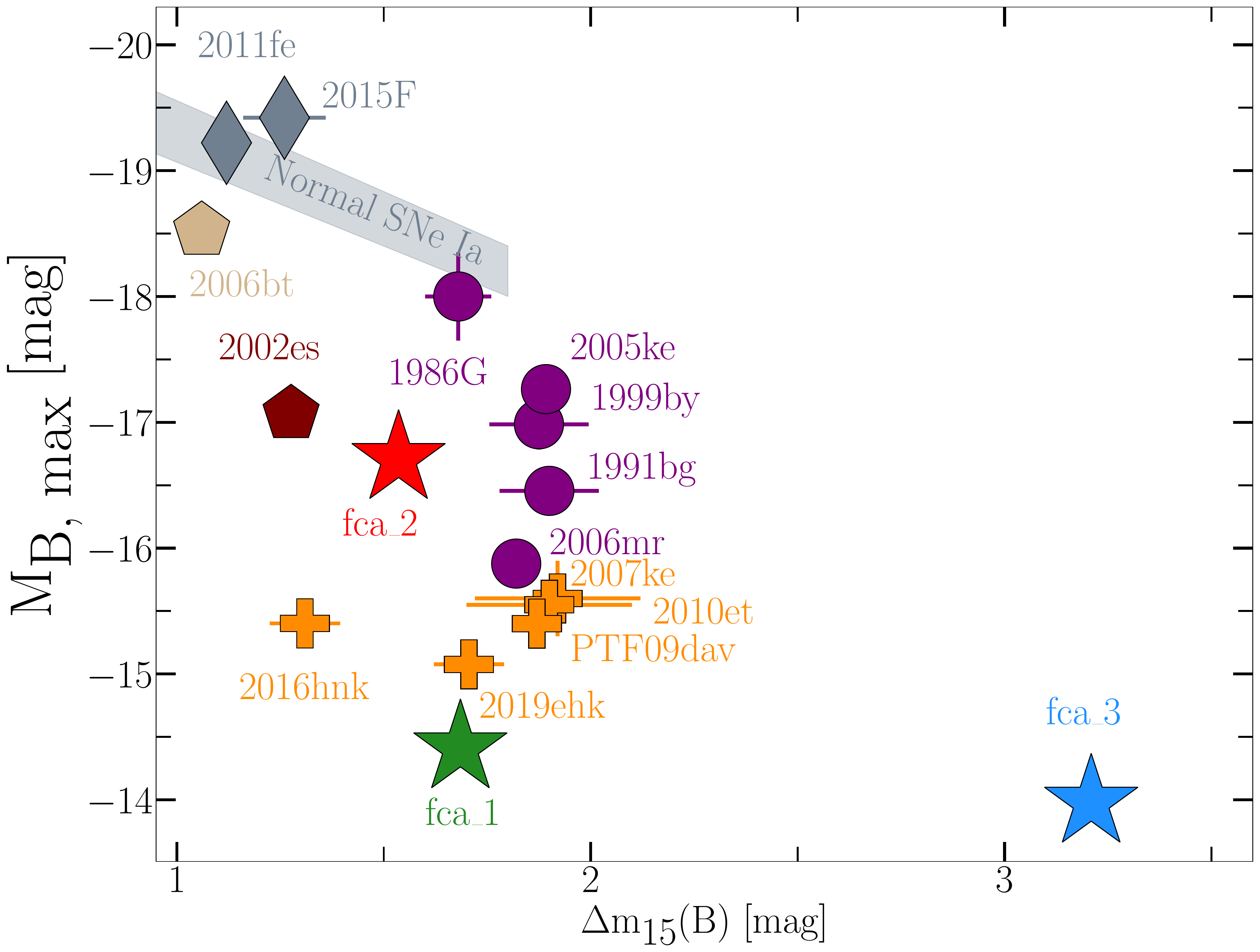}

\caption{Illustration of $\Delta m_{15}$ vs. $M_{\rm B,max}$ for models fca$_1$ (green star), fca$_2$ (red star) and fca$_3$  (blue star), and normal SNe Ia (diamonds+gray region), 91bg-like SNe Ia (circles), SNe Iax (stars), 02es-like SNe Ia (pentagons), other \cast{} (plus signs), and peculiar thermonuclear SN~2006bt (pentagon). Some uncertainties in $M_{\rm B,max}$ are smaller than the plotted marker size.
}
\label{fig:m15mb}
\end{figure*}

\begin{figure*}
\includegraphics[width=0.48\linewidth]{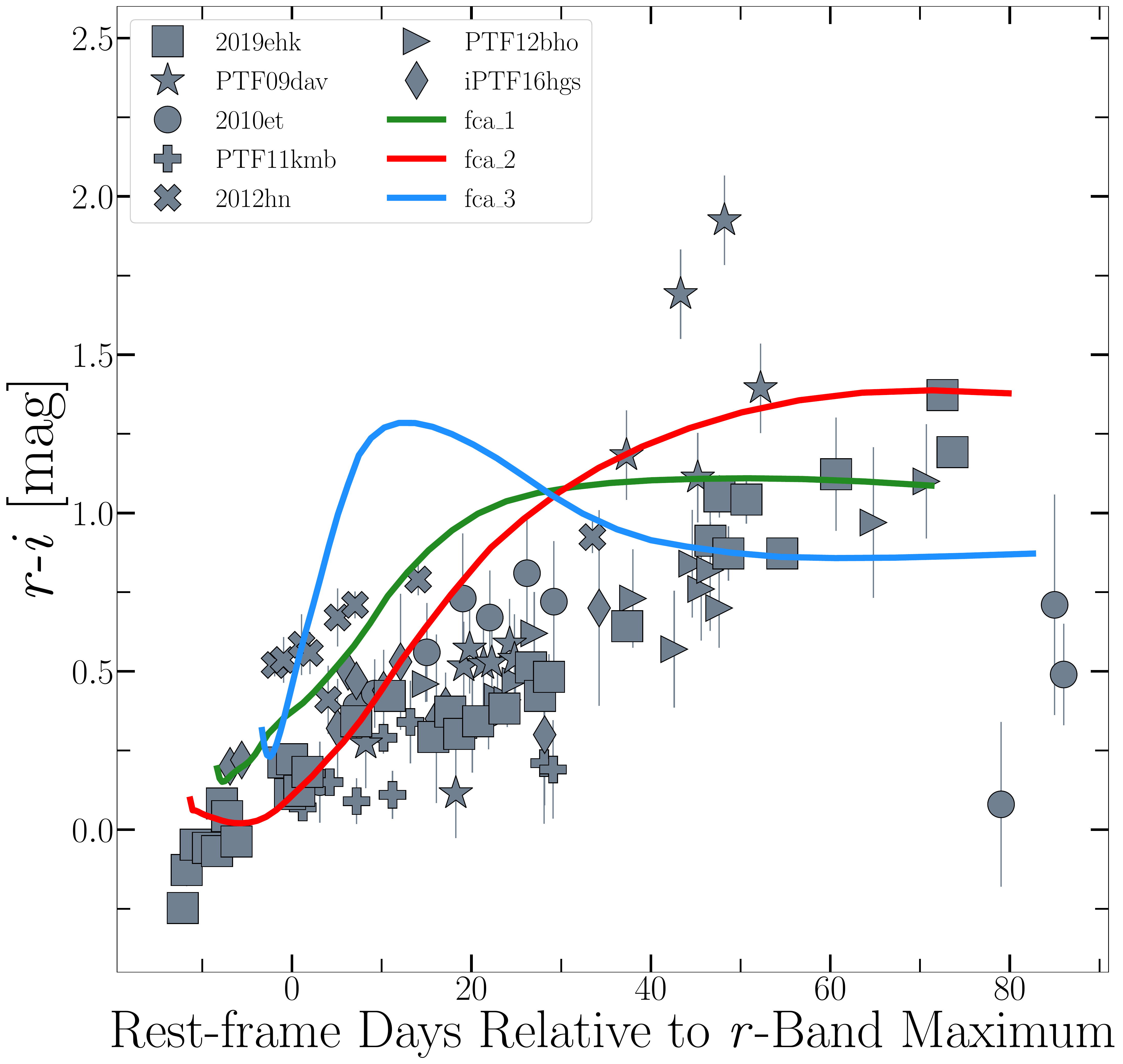}
\includegraphics[width=0.48\linewidth]{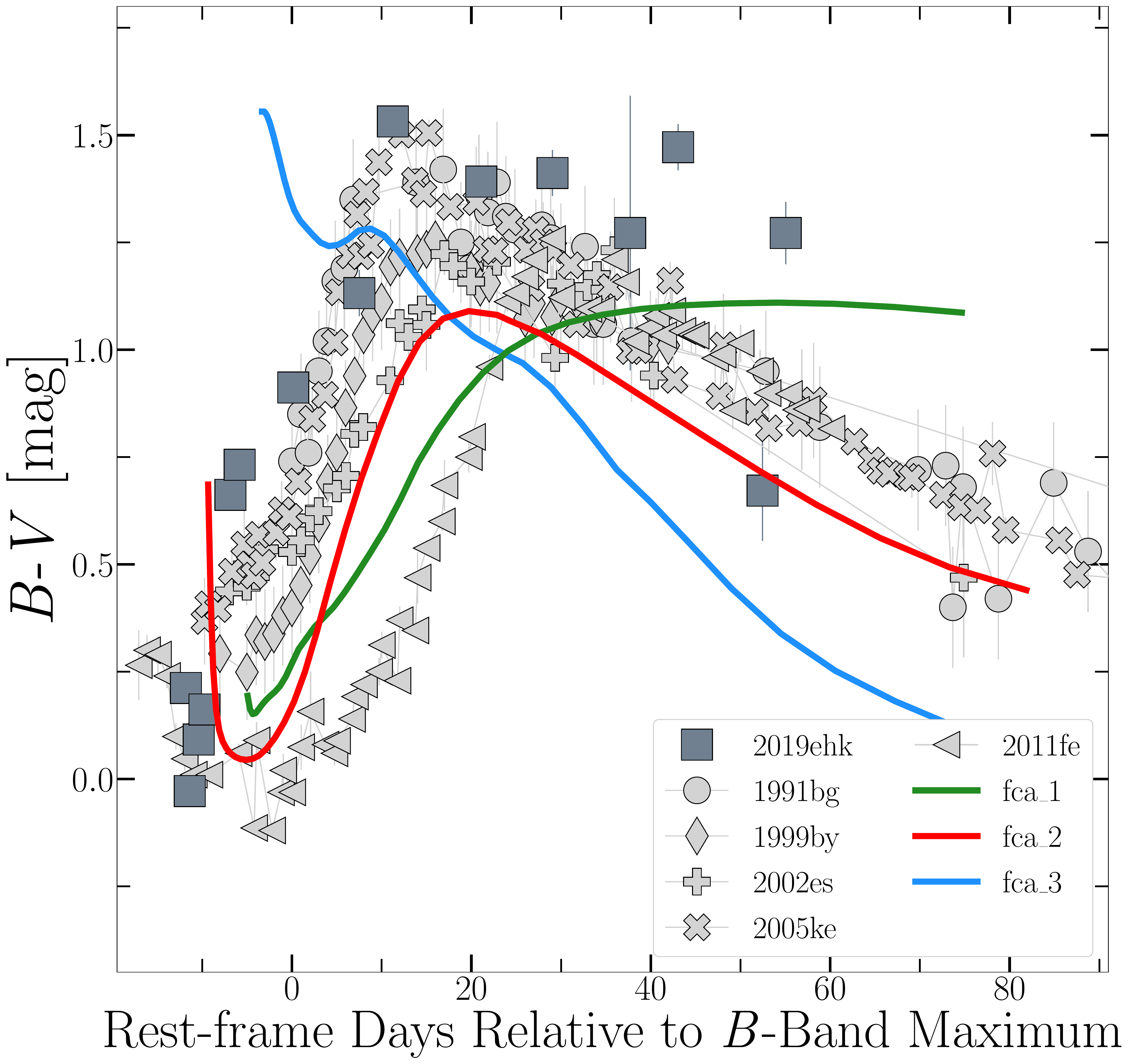}

\caption{Color evolution comparison of models fca$_1$ (green curve), fca$_2$ (blue curve) and fca$_3$  (red curve) and \cast{}. Left panel: r$-$i color comparison of models fca$_1$, fca$_2$ and fca$_3$  and \cast{}. Right Panel: B$-$V color comparison of the three models and various types of SNe Ia and SN2019ehk.}
\label{fig:dbands_ri_BV}
\end{figure*}

\section{Results}       
\label{Sec:Results}
\subsection{The disk evolution and thermonuclear explosion}
Our modelling begins with a debris disk assumed to have formed following the disruption of low-mass CO WD by a higher mass HeCO WD. 
As can be seen in Fig. \ref{fig:FLASH}, the disk evolves viscously, and the CO-WD debris is accreted onto the HeCO WD. The CO debris accreted onto the outer shell of the HeCO WD is then heated-up and compressed onto the He-shell of the HeCO WD, until sufficiently high temperatures and densities are reached and nuclear burning ensues. This is followed by a (weak) detonation in the He-rich material, producing nuclear burning products (see Fig. \ref{fig:flashSi28_Ni56}), and ejecting a large fraction of the burned material from the He shell of the HeCO WD and the debris disk. The explosion ejects a few times 0.1 M$_\odot$ of mostly burned material, composed of a significant fraction of intermediate elements and in particular 20-30$\%$ in Ca; and produces 0.01-0.05 M$_\odot$ of \niVI{} in the different models.  
We explore three pairs of WDs, and explored one of the models with a different viscosity parameter (but without using trace particles for the latter), for a total of four models, all showing a similar accretion and nuclear burning evolution, but show a range in the specific timescales of the evolution, and the overall synthesis of nuclear burning products. The model properties are listed in Table \ref{tab:models}. The resulting ejecta, bound material and energetics are presented in Table \ref{tab:results}. The compositions of the ejecta in the different models are summarized in Table \ref{tab_composition}.

\subsection{Light curves and spectra}
Using the non-LTE radiation transfer code (\texttt{CMFGEN}), we produce the light curves and spectra resulting from each of the models. The \textit{UBVIRJHK} LCs are shown in Fig. \ref{fig:Allbands}. As can be seen in Fig. \ref{fig:Rbandlc} and \ref{fig:lcBol_IIb}, the LCs  compare well with the range of observed \cas{} SNe (not including the shock breakout early peak, likely due to CSM interactions, \citealt{WynnV+20b}, also seen in X-rays, not modeled in the current simulations). These are also compared to various other types of SNe in Fig. \ref{fig:lcBol_IIb}. Fig. \ref{fig:dbands_ri_BV}\ shows the color evolution of the models in comparison with various Ia SNe and the \cas{} SN SN 2019ehk.
Fig. \ref{fig:spectraNLTE} shows the good comparison of the spectra with that of \cas{} SNe (the detailed evolution of the modeled spectra can be seen in the appendix). Fig. \ref{fig:m15mb} shows the positions of the modeled SNe in the peak-width relation in comparison with various types of SNe, showing the small differences in the WD masses can span a wide range of behaviour, in terms of peak luminosity and the rate of LC evolution . A detailed discussion of the comparison between the modeled LCs and spectra and the observation can be found in the discussion section below.  

 Overall, our resulting models show excellent agreement with the observed \cas{} SNe and their range of properties (besides the lack of clear He lines, which are observed in many of the \cas{} SNe; this issue is discussed below). In fact, to the best of our knowledge, these are the first models to well reproduce both light curve and spectral evolution of \cas{} SN, as well as reproduce the range of behaviors (compare with e.g. \citealt{Waldman+11} and \citealt{D15_hedet}).  
  We should also note that the specific models were not chosen as to fit any of the observations, but rather we present all the models we ran to date.
 
\subsection{Ejecta composition and line strength}
As can be seen in tables \ref{tab:results} and \ref{tab:IsoModels}, the resulting SNe produce large abundances of intermediate elements and low \niVI{} mass, consistent with the inefficient burning of He-rich material. In particular large fractions of the ejecta mass are composed of Ca, which give rise to strong Ca lines.
As mentioned above, it was suggested that the Ca abundances might not need to be high in \cas{} SN, but rather that the high nebular line ratios of Ca to O in these SNe is due to Ca being such an effective coolant, and that these objects are just O-poor \citep{Shen+19}.
In our models, however, the Ca abundances are inherently very high, and our successful models are indeed \cas{} SNe, rather than just having strong Ca lines relative to O.

We do note in regard to the OI6300 line strength could be sensitive to the relative fractions of Ca and O in different regions of the ejecta. In particular the offset in abundance between Ca and O can be very large in different regions, when the material is not well mixed \citep[see][]{D15_hedet}. Since the cooling power of Ca only affects the regions where it is present, different mixing of Ca and O can potentially make large differences in the large ratios produced. 
In our study we made use of a 1D non-LTE code in order to model the spectra resulting from the radiative transfer analysis, and therefore needed to map the 2D simulation data into 1D, as discussed above, potentially leading to some artificial mixing and composition smoothing. We therefore should note the caveat that stronger OI6300 line might be have been produced without such artificial mixing.

\section{Discussion}  \label{Sec:discussion}
Our models of the disruptions of low-mass CO WDs by higher mass hybrid HeCO WDs produce faint, \cas{} SNe which agree well with the observed class of \cas{} SNe \citep{Perets+10}, and can reproduce the range of brighter, slower-evolving \cas{} SNe down to the fainter and faster-evolving \cas{} SNe. In the following we discuss in detail the comparison of the models with the observations, the various possible caveats in the models and the overall implications of our results. We also briefly discuss the demographics of \cas{} SNe in this context, but postpone a more detailed discussion of the latter to a dedicated follow-up paper.

\subsection{Comparison to \cast{} Observations}  \label{subsec:Obs} 

The fca$_1$, fca$_2$ and fca$_3$ models contain many attributes that make them consistent with the observed properties of \cast{}. In Figure \ref{fig:m15mb} we present the Phillips relation in the B-band for a variety of SNe Ia, as well as some \cast{}. While both the fca$_1$ and fca$_2$ models are consistent with the decline rate ($\Delta m_{15}(B)$) for \cast{} 2007ke, PTF09dav, 2007ke, 2016hnk, and 2019ehk, the former produces a slightly fainter B-band maximum and the latter is too luminous compared to observations. Furthermore, the fca$_3$ model declines much faster than these \cast{} and its maximum B-band light curve absolute magnitude is fainter, which could be consistent with some of the lowest luminosity \cas{} events. Overall, the r$-$i colors (Fig. \ref{fig:dbands_ri_BV}) of the fca$_1$ and fca$_2$ models are consistent with the observed color evolution of \cast{}; these models also shows some consistency to the B$-$V colors of SNe Ia varieties shown in Figure \ref{fig:dbands_ri_BV} but are slightly bluer overall than SN~2019ehk's evolution. 

In Figure \ref{fig:Rbandlc} we compare the $r-$band light curves of 10 \cast{} to fca$_1$, fca$_2$ and fca$_3$ models. The fca$_1$ model provides a nearly identical match to the light curve of SN~2019ehk after the primary light curve peak, whose physical origin (e.g., circumstellar interaction or shock cooling emission) is unexplored by these models. While the evolution of the fca$_2$ r-band light curve matches that of observed \cast{}, it $\sim$1-2 magnitudes brighter than most of the sample. However, the fca$_2$ shows a remarkable consistency with the peak magnitude and decline rate of Ca-strong SN~2016hnk, which was thought to be either a 1991bg-like SN~Ia \citep{galbany19} or the result of a He-shell detonation of a sub-Chandrasekhar mass WD \citep{wjg19}. Furthermore, despite its fast rise time relative to \cast{}, the fca$_3$ model matches the r-band light curve peak and decline rate of a number of objects such as iPTF16hgs, SN~2005E, PTF11kmb and SN~2021gno. The double-peaked light curve structure is not produced in either model, however, this is expected given that this phenomenon is thought to arise with shock cooling emission or shock interaction of extended circumstellar material, as observed in iPTF16hgs, SNe~2019ehk, 2021gno and 2021inl. 

Additionally, fca$_2$ and fca$_3$ models appear inconsistent with observations of SNe~IIb/Ib (Fig. \ref{fig:lcBol_IIb}; right). In terms of bolometric luminosity (Fig. \ref{fig:lcBol_IIb}; left), the fca$_1$ and fca$_2$ models are consistent with the evolution of SN~2019ehk during and after its secondary light curve peak, while the fca$_3$ model is only consistent with SN~2021gno's peak luminosity but not its slower decline rate.

In Figure \ref{fig:spectraNLTE}, we present the early- and mid-time spectra of all three models relative to observed \cas{} SNe. Near light curve maximum, we find that the fca$_1$ model shows consistent spectral features of Ca II, He I and Si II to \cast{} and even reproduces the early-time emergence of the nebular [Ca II] transition. Furthermore, the fca$_1$ model spectrum is very similar to SNe~2005E, 2007ke, and 2019ehk at phases $\sim 20-60$~days post-maximum. At this point in their evolution, both the fca$_1$ model and observed \cast{} spectra are dominated by Ca II emission as well as a weak [O I] emission profile, which is in line with the suggested observational definition of Ca-strong SNe  that the line ratio [Ca II]/[O I] is greater than 2. Overall, the consistency of this explosion model indicates that its ejecta structure is likely very similar to that of many observed \cast{}. 

\subsection{He abundances}
Our HeCO WDs are initially composed of a non-negligible He mass in the WD outer shell. However, in our current models these large He abundances exiting in the initial HeCOs are mostly burned, leaving only small abundances of He in the ejecta. In particular, the He contribution is too small as to leave strong He spectral signatures. In that sense, our models can well reproduce He-poor \cas{} SNe, but producing \cas{} SNe with strong He line could be more challenging (though one should note that good and very secure identification of He-lines in the observed \cas{} SNe is complex; see e.g. \citealt{De+20}). That being said, here we only considered a few models of HeCO WDs disrupting a CO WDs. Other branches of HeCO WDs could give rise to larger He abundances (see e.g. our discussion in \citealt{Pakmor+21}), and in other cases two HeCO WDs could merge (as we find in population synthesis modelling; to be discussed in a later paper). In both these cases larger He abundances are available during the disruption and explosion, potentially leaving behind much larger He abundances in the ejecta. In particular, in a preliminary model where we studied the disruption of a HeCO WD by another HeCO WD (not shown here) using similar methods , we found that the ejecta has similar IGE, has somewhat lower IME abundances,  but has far larger He abundances in the ejecta ($\rm M(He)\gtrsim 7 \times 10^{-2}\Msun{}$ and $\rm M($\caIV$) \sim 0.08 - 0.12 \Msun$). This model and other more He-rich models will be discussed in a dedicated paper (Zenati, Perets et al., in prep). It is therefore possible the the He-strong \cas{} SNe arise from a very similar channel, but in more He rich mergers due to double HeCO WDs or more He-rich HeCO WDs.

\subsection{High production of Ca \TiIV{} and the contribution of positions to the Galactic 511 kev emission} 
\TiIV{} is produced from the fusion of an $\alpha$- nuclei with \caIV{} atoms. Excessive production of \TiIV{} is therefore typically accompanied by the excessive production of other intermediate $\alpha$ elements. Highly fine-tuned conditions are therefore required in order to produce large abundances of one of these isotopes without producing significant abundances of the other. Indeed, the fractional ratio of \TiIV{} to stable Calcium, \caIV{} is found to be large in all studied cases of Helium-detonations. Already the original identification and characterization of \cas{} SNe as thermonuclear SNe, inferred a large abundance of Ca, with an estimated mass of $\sim$ 0.1 $M_{\odot}$ of Calcium in SN 2005E, the prototype SN for this type of SNe. \cas{} SNe are therefore likely candidates as significant production sites of \TiIV{}.  By taking the range of \TiIV{} to \caIV{} ratios found in theoretical models and the estimated Calcium mass in SN 2005E one can infer that large \TiIV{} abundances should be produced.

While some of the previous models have indeed shown a large production of Ca \citep{Waldman+11}, they did not well reproduce the observed light curves. The simulations shown here do provide a potentially successful model for \cas{} SNe. The composition of the ejecta from these models is shown in Table \ref{tab:IsoModels}. As can be seen, Ca is indeed excessively produced in these models and as much as 20-30 $\%$ of the ejecta is composed of Ca, supporting the identification of 2005E-like SNe as \cas{} SNe, rather than SNe with strong Ca lines, as mentioned above. Furthermore, our models show a significant production of \TiIV{} with as much as 1-8$\times10^{-2}$ M$_\odot$ of ejected \TiIV{}. In \citep{Perets+10,Perets14} we suggested that the positrons produced through the later radioactive decay of \TiIV{} in \cas{} SNe could provide a major contribution to the Galactic 511 kev emission, which origins are still not understood. \cite{Perets14} assumed the \TiIV{} fractions produced in \cas{} SNe followed the results in \citep{Waldman+11} models, which were 1-2 orders of magnitude smaller than found in our current models. 
The even larger \TiIV{} abundances found here suggest a potentially even more important contribution of \cas{} SNe to 511 kev production, even if the rate of \cas{} SNe in the Galactic bulge smaller than those assumed in \citep{Perets14}.

The large abundances of Ca initially inferred from observations \citep{Perets+10} were suggested to play an important role in explaining the origins of enhanced Ca abundances in the intracluster medium \citep{Perets+10,Mul+14}. The possibility that these SNe are only Ca strong-lined rather than inherently Ca rich \citep{Shen+19} could potentially exclude this Ca enrichment scenario. However, our current results do support the large Ca production, and hence their potential role in explaining the composition of the intracluster medium.     

\subsection{Polarization}
Our radiative transfer calculations do not model polarization, and we therefore can not directly provide any predictions in this regard. Nevertheless, we point out that our models produce non spheri-symmetric explosions, unlike our \cite{Perets+19}, and others' models for normal type Ia SNe which are typically quite spherical (see, \citealt{Soker19} table 1). Polarization measurements of type Ia SNe show them not to be polarized (or very weakly polarized) consistent with the various suggested models. We hypothesize that asphericity of our models for \cas{} SNe would give rise to polarized radiation. Polarization measurement are typically quite difficult and require significant fluxes, but future polarization measurements of sufficiently close-by \cas{} SNe might be possible, and could test this prediction, and, if correct, show them to be far more polarized than normal type Ia SNe.   

\subsection{Possible caveats}
While potentially providing the first  successful model for \cas{} SNe, one needs to consider several potential caveats.

\begin{itemize}
    \item Our models are axisymmetric 2D models, and in particular do not consider the early disruption stage. Our disk models are therefore not directly derived from a self-consistent modeling of the disk formation, which might differ from the hypothesised disk used in our models.
    \item Our radiative transfer modeling is 1D, while the explosion is likely not spheri-symmetric, and a 2D, or better 3D radiative transfer modeling might be required in order to better represent the predicted light-curve and spectra.
    \item The viscous evolution of our disks is modeled simplistically through artificial Shakura-Sunayev viscosity, and a more realistic evolution could potentially affect the results. 
    \item While our models provide good match for \cas{} SNe, they poorly He-lines compare to the observed for many of the \cas{} SNe, suggesting more He should exist in the ejecta (though as we briefly discuss this issue might be alleviated with HeCO-HeCO WD mergers of more He-enriched HeCO WDs).
    
\end{itemize}

Future 3D modeling of mergers (currently in progress) and 3D radiative transfer models might resolve the first two issues, and extending our models to double HeCO WDs (in progress) and more He-enriched HeCO WDs would allow us to test the possibility of producing more He-rich \cas{} SNe.

In regards to the viscosity,
we have run one of the models with a different viscosity parameter, as to get a handle of its influence (see Table \ref{tab:models}). The resulting evolution, explosion energetics and composition, while somewhat quantitatively different than the original model, do show a similar qualitative behaviour, providing support to the robustness of the results in this regard. Nevertheless, more realistic simulations of this issue could better test this.

\begin{table*}
\begin{centering}
\begin{tabular}{|c|c|c|c|c|c|c|c|c|}
\hline 
\# & ${\rm fca1_{unbound}}$  & ${\rm fca2_{unbound}}$  & ${\rm fca3_{unbound}}$ & ${\rm fca1_{bound}}$  & ${\rm fca2_{bound}}$  & ${\rm fca3_{bound}}$ \tabularnewline

- & $[\rm M_{\odot}]$ & $[\rm M_{\odot}]$ & $[\rm M_{\odot}]$ & $[\rm M_{\odot}]$ & $[\rm M_{\odot}]$ & $[\rm M_{\odot}]$\tabularnewline
\hline 
\hline 
$^{1}$H & $1.31\times 10^{-5}$  & $4.18\times 10^{-5}$  & $6.05\times 10^{-5}$ & $2.77\times 10^{-5}$  & $1.69\times 10^{-5}$  & $2.38\times 10^{-5}$ \tabularnewline

$^{4}$He & $2.55\times 10^{-3}$  & $9.27\times 10^{-4}$  & $8.35\times 10^{-4}$ & $8.47\times 10^{-3}$  & $4.83\times 10^{-3}$  & $3.77\times 10^{-3}$ \tabularnewline

$^{12}$C & $1.01\times 10^{-1}$  & $1.04\times 10^{-1}$  & $8.84\times 10^{-2}$ & $4.86\times 10^{-2}$  & $5.39\times 10^{-2}$  & $5.62\times 10^{-2}$ \tabularnewline

$^{14}$N & $1.69\times 10^{-7}$  & $3.62\times 10^{-7}$  & $1.68\times 10^{-6}$ & $3.46\times 10^{-7}$  & $1.45\times 10^{-7}$  & $5.32\times 10^{-7}$ \tabularnewline

$^{16}$O & $9.82\times 10^{-2}$  & $9.91\times 10^{-1}$  & $1.55\times 10^{-1}$ & $4.56\times 10^{-2}$  & $2.97\times 10^{-2}$  & $4.22\times 10^{-2}$ \tabularnewline

$^{20}$Ne & $8.24\times 10^{-3}$  & $1.12\times 10^{-2}$  & $1.19\times 10^{-2}$ & $3.28\times 10^{-5}$  & $3.42\times 10^{-4}$  & $8.48\times 10^{-5}$ \tabularnewline

$^{24}$Mg & $3.17\times 10^{-3}$  & $1.83\times 10^{-5}$  & $5.08\times 10^{-5}$ & $7.63\times 10^{-5}$  & $6.33\times 10^{-6}$  & $3.26\times 10^{-6}$ \tabularnewline
 
$^{28}$Si & $2.28\times 10^{-3}$  & $3.31\times 10^{-2}$  & $8.89\times 10^{-4}$ & $1.31\times 10^{-3}$  & $1.07\times 10^{-3}$  & $2.08\times 10^{-3}$ \tabularnewline
 
$^{32}$S & $8.23\times 10^{-3}$  & $6.44\times 10^{-3}$  & $2.65\times 10^{-3}$ & $2.46\times 10^{-4}$  & $2.61\times 10^{-4}$  & $1.12\times 10^{-4}$ \tabularnewline

$^{35}$Cl & $3.19\times 10^{-4}$  & $3.05\times 10^{-4}$  & $8.51\times 10^{-4}$ & $1.62\times 10^{-5}$  & $2.51\times 10^{-5}$  & $1.96\times 10^{-5}$ \tabularnewline

$^{36}$Ar & $6.57\times 10^{-4}$  & $7.92\times 10^{-4}$  & $1.99\times 10^{-4}$ & $7.32\times 10^{-4}$  & $1.42\times 10^{-4}$  & $2.15\times 10^{-5}$ \tabularnewline

$^{40}$Ca & $1.37\times 10^{-1}$  & $1.08\times 10^{-1}$  & $1.34\times 10^{-1}$ & $3.06\times 10^{-3}$  & $1.58\times 10^{-3}$  & $8.16\times 10^{-4}$ \tabularnewline

$^{44}$Ti & $7.94\times 10^{-3}$  & $1.46\times 10^{-2}$  & $7.34\times 10^{-3}$ & $2.41\times 10^{-4}$  & $6.03\times 10^{-4}$  & $2.15\times 10^{-4}$ \tabularnewline

$^{48}$Cr & $1.02\times 10^{-5}$  & $1.94\times 10^{-6}$  & $3.24\times 10^{-4}$ & $4.02\times 10^{-6}$  & $1.06\times 10^{-6}$  & $4.52\times 10^{-6}$ \tabularnewline

$^{52}$Fe & $2.44\times 10^{-5}$  & $7.03\times 10^{-5}$  & $5.51\times 10^{-4}$  & $1.16\times 10^{-7}$  & $2.82\times 10^{-6}$  & $4.35\times 10^{-5}$ \tabularnewline

$^{54}$Fe & $5.29\times 10^{-4}$  & $1.19\times 10^{-3}$  & $1.54\times 10^{-3}$  & $1.66\times 10^{-5}$  & $4.59\times 10^{-6}$  & $2.13\times 10^{-5}$ \tabularnewline

\niVI{} & $2.12\times 10^{-2}$  & $5.05\times 10^{-2}$  & $1.06\times 10^{-2}$ & $5.33\times 10^{-5}$  & $7.12\times 10^{-5}$  & $6.02\times 10^{-5}$ \tabularnewline
\hline 
\end{tabular}
\par\end{centering}
\caption{\label{tab:X-ejectamodels}The unbound and bound material of the ejecta mass of all models. \protect \\ }
\label{tab:IsoModels}
\end{table*}

\section{\cas{} SNe demographics and environments}   \label{Sec:PPS}
\cas{} SNe are observed both in early type and late type galaxies, but they are far more abundant in early type galaxies compared with normal type Ia SNe \citep[e.g.][]{Perets+10,Kasliwal+12,Lym+13,De+20}, and typically explode far from star-forming regions, suggesting a much older progenitor population for these SNe \citep{Perets+10,Perets+11,Lym+13}. Such SNe are also observed in relatively large offsets from their host galaxy nuclei when observed in early type galaxies \citep{Perets+10,Perets14,Foley+15,PeretsBeniamini21}, consistent with the existence of large old stellar populations at the large halos of early type galaxies \citep{PeretsBeniamini21}. 
Together, these suggest an overall a delay time time distribution far more skewed towards long Gyrs delays compared with normal type Ia SNe. Assuming hybrid HeCO WDs disrupting lower-mass WDs are the progenitors of \cas{} SNe, as suggested here, one can explore the rates of such SNe and their delay time distribution using population synthesis studies. The data in our original population synthesis study in \citep{Perets+19} included information on such progenitors; analyzing these data we found the such progenitors can produce rates as large as a few up to 15$\%$ of the normal type Ia rate, and can have very long delays, far more extended than normal type Ia SNe. These are potentially consistent with the observed rates and demographics of \cas{} SNe. However, these could be sensitive to the exact choice of the mass limits of the HeCO and CO WD progenitors, the metallicities etc. We therefore postpone the detailed exploration of these issues to a dedicated paper (Toonen et al., in prep.).  

\section{Summary}     
\label{Sec:conclusion}
When we first characterized \cas{} SNe as belonging to a unique novel type of faint, Ca and He-rich SNe, we found many of them explode in very old environments \citep{Perets+10}, and inferred them to have low mass ejecta. These findings already excluded a core-collapse origin for at the majority of these SNe, and pointed to a likely thermonuclear origin, arising from a He-rich explosion of a WD. Extensive observational studies of these SNe over the last decade far improved and extended our understanding of their properties.  It showed them to be more diverse than originally thought \citep{Kasliwal+12,De+20}, and include He-poor (typa Ia/c) SNe, and the existence of early CSM interactions \citep{WynnV+20b}. Less progress was done on the theoretical understanding of these SNe. 

Decades ago \citep{Woo+86} found that that detonation of large He-shells on low-mass WDs (suggested to form following long-term mass-transfer onto a WD from a stellar companion) might produce \niVI{} poor, and possibly \cas{} SNe. These were not likely to resemble normal type Ia SNe, and, consequently, they attracted little interest and follow-up studies at the time (although they were suggested to be related to rapidly evolving SNe 1885A and 1939B; \citealt{deV+85,per+11b}). 

The finding and characterization of a novel type of faint, \cas{} SNe \citep{Perets+10} renewed the interest in such He-shell detonation models, together with more modern incarnations of the \citep{Woo+86}, termed .Ia SNe \citep{Bildsten+07} which considered large He-shell detonation models. These models, and new ones that produced for the first time detailed light curve and spectra of such scenarios, while having some success in producing the observed and/or inferred properties of \cas{} SNe, did not well reproduce the overall properties \citep{Waldman+11,D15_hedet}. Suggestions of NS mergers with WDs were also explored as possible progenitors \citep{Fernandez&Metzger13}, but more detailed studies of such explosions excluded these as potential progenitors for \cas{} SNe \citep{Zenati19b,Toonen+18}. 

Here we proposed a novel model for the origin of \cas{} SNe from the disruption of low-mass CO WDs by hybrid HeCO WDs, and the subsequent accretion of the CO-WD debris on the HeCO WD. We made use of 2D hydrodynamical-thermonuclear simulations of the evolution of the debris disk, and showed a weak He-rich detonation ensues following the accretion of CO material on the HeCO He-shell (surface). This weak explosion ejects a few 0.1 M$_\odot$ of intermediate-elements rich material and produce up to a few 0.01 M$_\odot$ of \niVI{}. Using nucleosynthesis post-processing and non-LTE radiation transport simulations, we modeled the light-curve and spectra of such explosions, and showed them to well match the light curve and spectra of \cas{} SNe. 

We modeled three different HeCO-WD CO-WD pairs, which produce a range of \cas{} explosive transients, providing the best models to date for the light curves and spectra of \cas{} SNe, and potentially explaining the range of observed \cas{} SNe (those not showing clear He lines). We suggest that the models of double HeCO WD mergers and/or more enriched HeCO WDs might also produce \cas{} SNe with more pronounced He lines.

\section*{Acknowledgements}
YZ thanks Kevin Schlaufman and Colin Norman for helpful discussions. YZ and HBP acknowledge support for this project from the European Union's Horizon 2020 research and innovation program under grant agreement No 865932-ERC-SNeX. W.J-G is supported by the National Science Foundation Graduate Research Fellowship Program under Grant No.~DGE-1842165. W.J-G acknowledges support through NASA grants in support of {\it Hubble Space Telescope} programs GO-16075 and 16500. \flash{} simulations were performed on the Astric computer cluster of the Israeli I-CORE center. The plots in this paper have been generated using \texttt{Matplotlib} \citep{Hunter07} and yt-project \citep{Turk+11}.

%\software{}

\bibliography{references}
\bibliographystyle{apj}
\bibliographystyle{aasjournal} 

%%%%%%%%%%%%%%%%%%%%%%%%%%%%%%%%%%%%%%%%%%%%%%%%%%%%%%%%%%%%%%%%%%%%%%%%%%%%%%%%%%%
\clearpage
\appendix

\begin{figure*}
\includegraphics[width=0.48\linewidth]{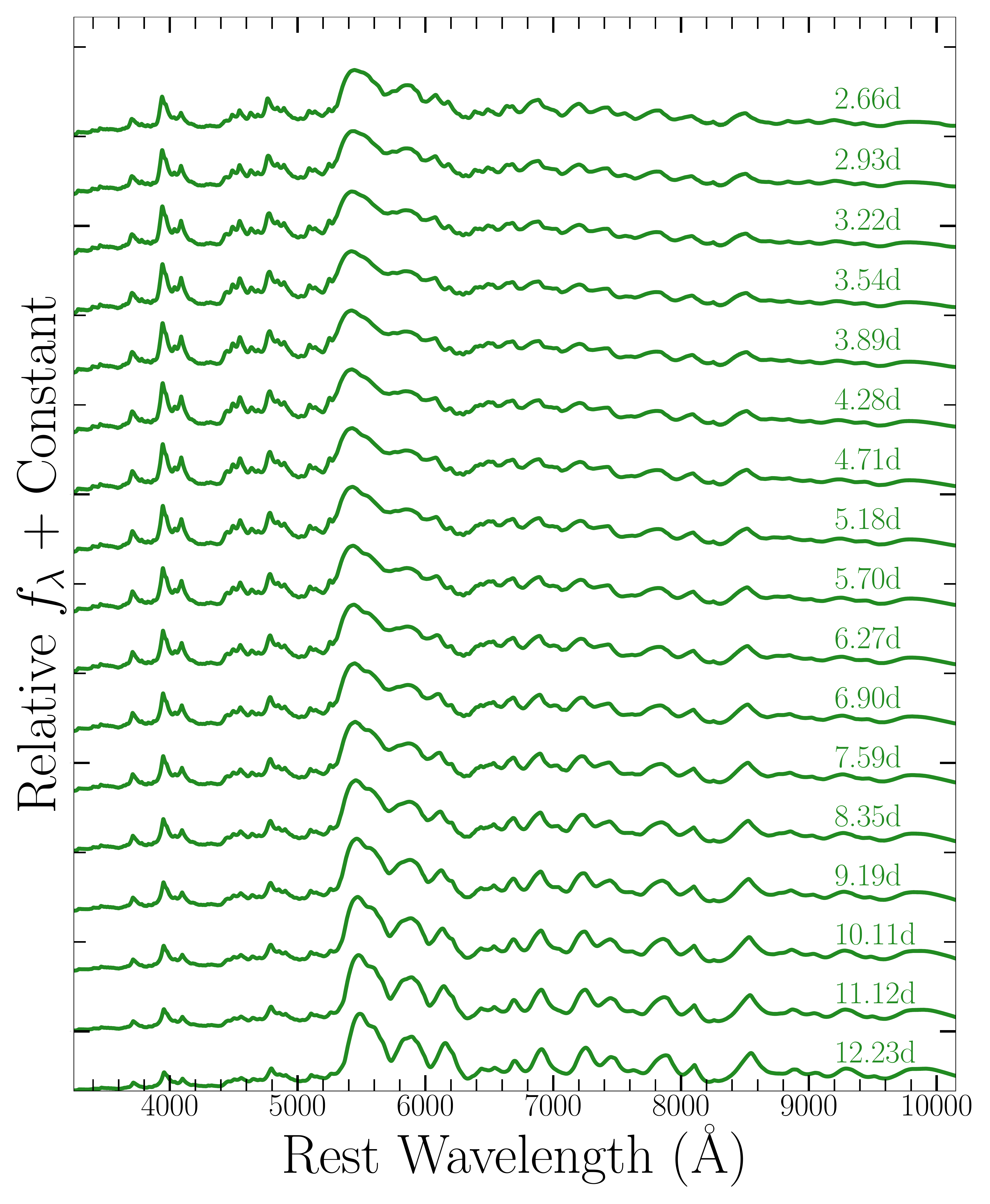}
\includegraphics[width=0.48\linewidth]{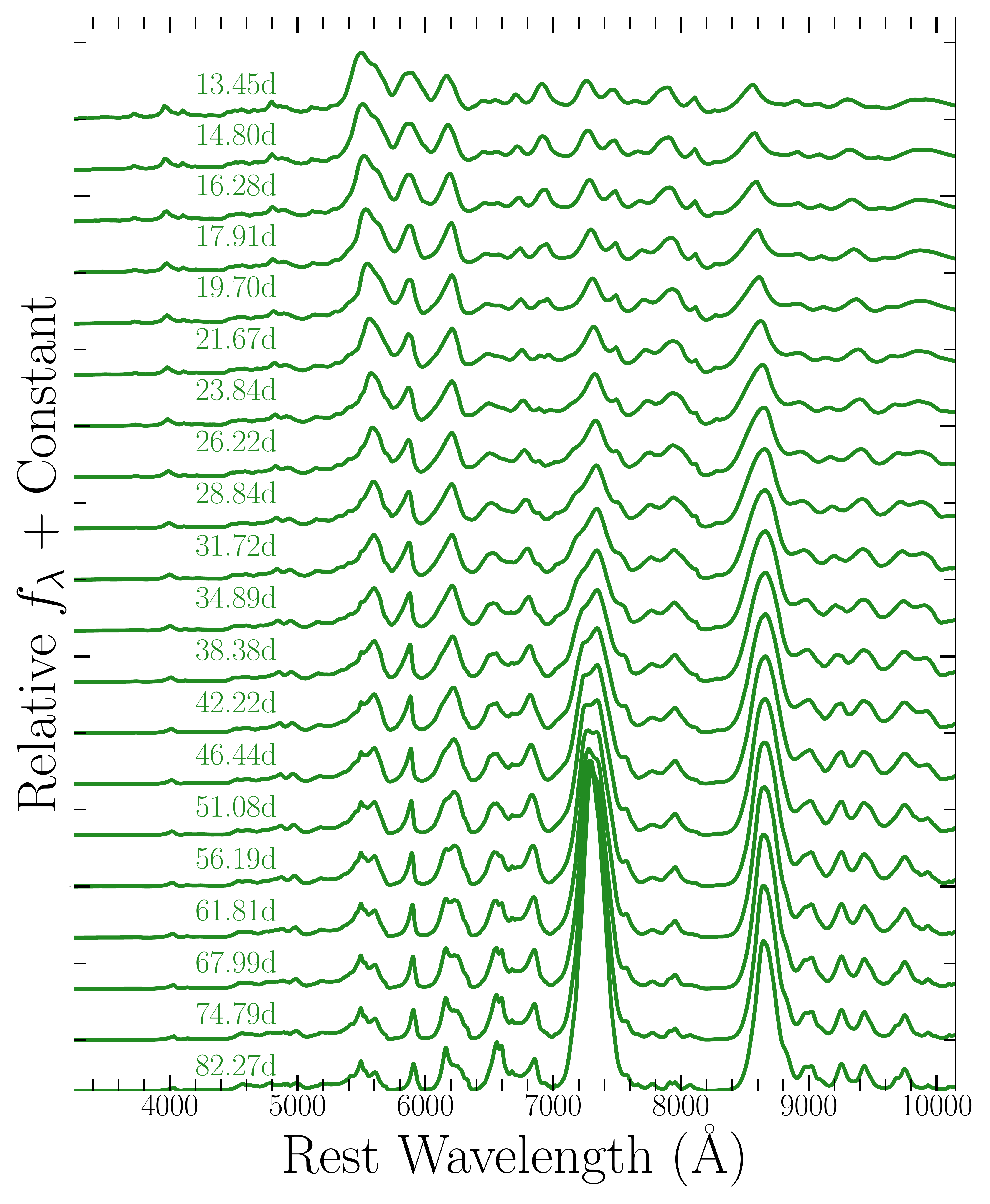}

\caption{Spectral series of fca$_1$ model. Each spectrum is normalized by the mean spectral flux.}
\label{fig:spectra_fca2}
\end{figure*}

\begin{figure*}
\includegraphics[width=0.48\linewidth]{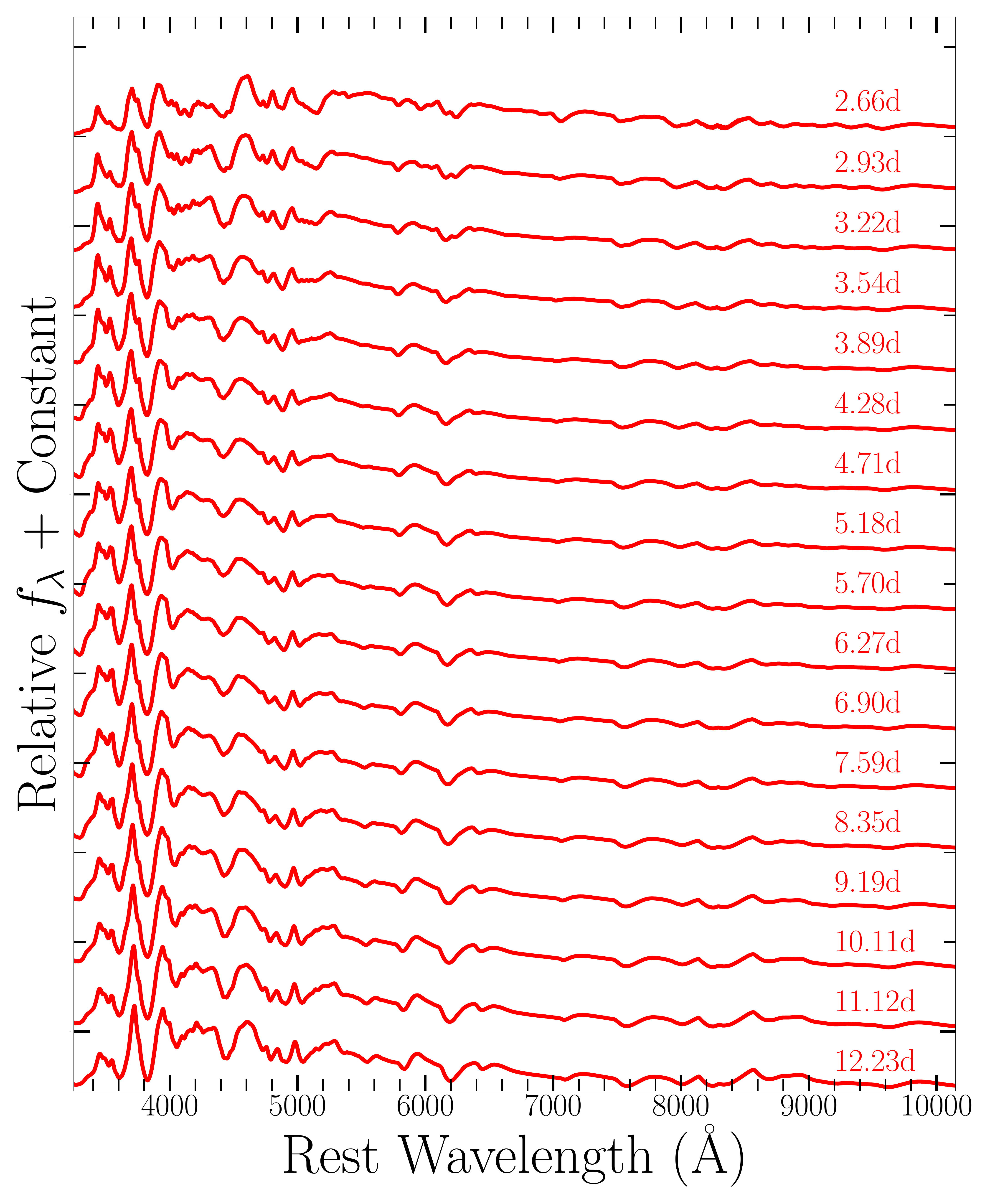}
\includegraphics[width=0.48\linewidth]{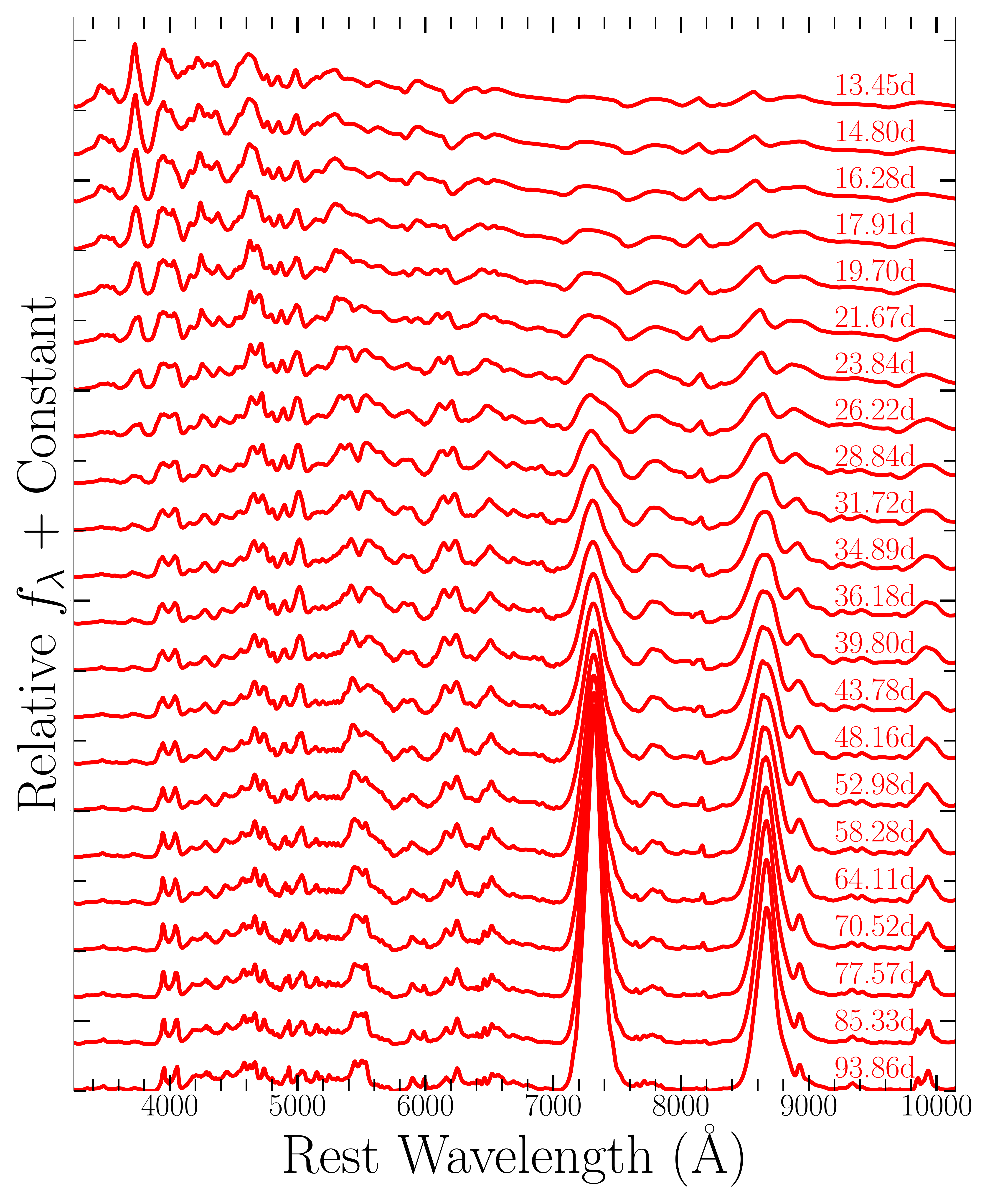}

\caption{Spectral series of fca$_2$ model. Each spectrum is normalized by the mean spectral flux.}
\label{fig:spectra_fca2}
\end{figure*}

\begin{figure*}
\includegraphics[width=0.48\linewidth]{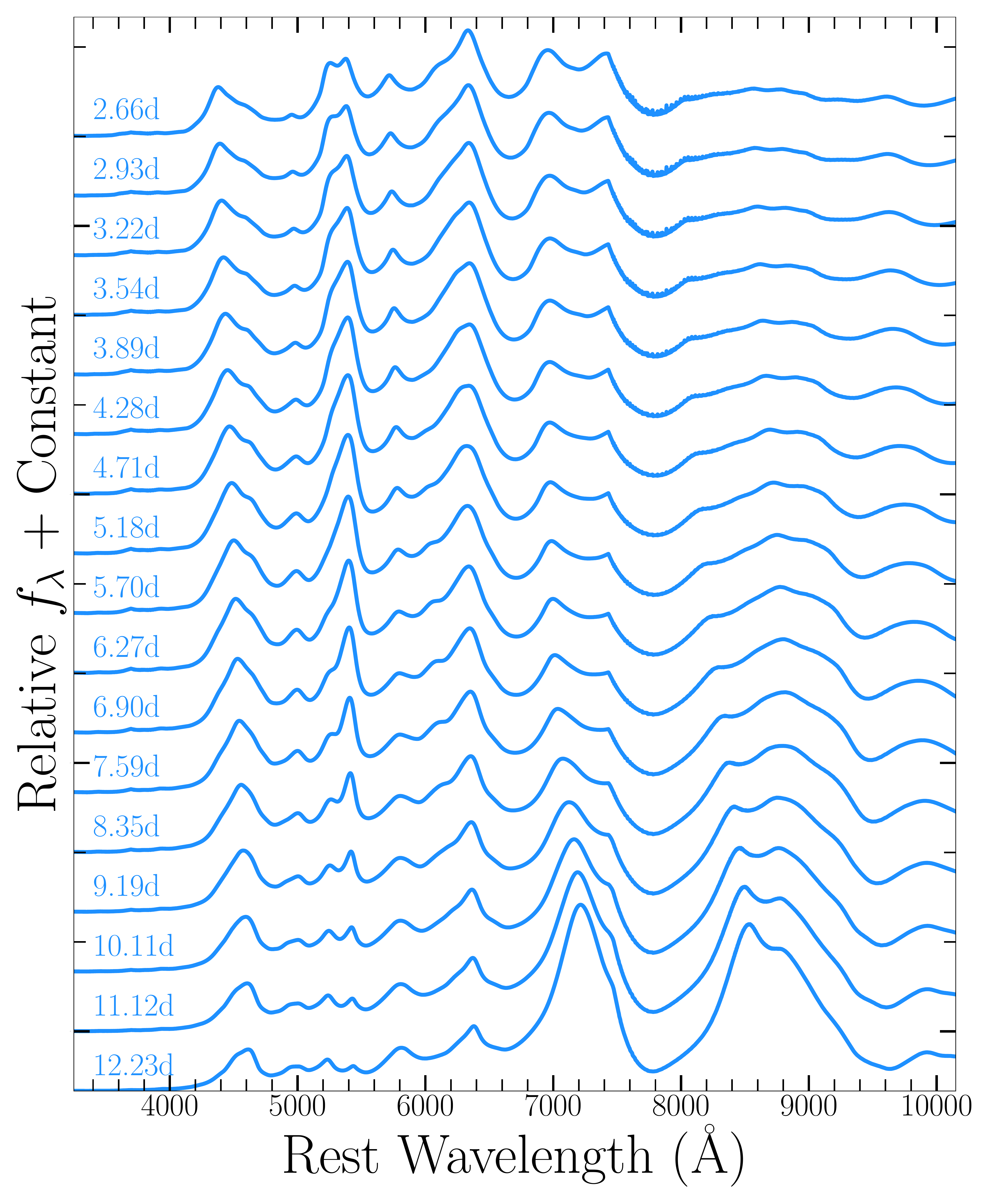}
\includegraphics[width=0.48\linewidth]{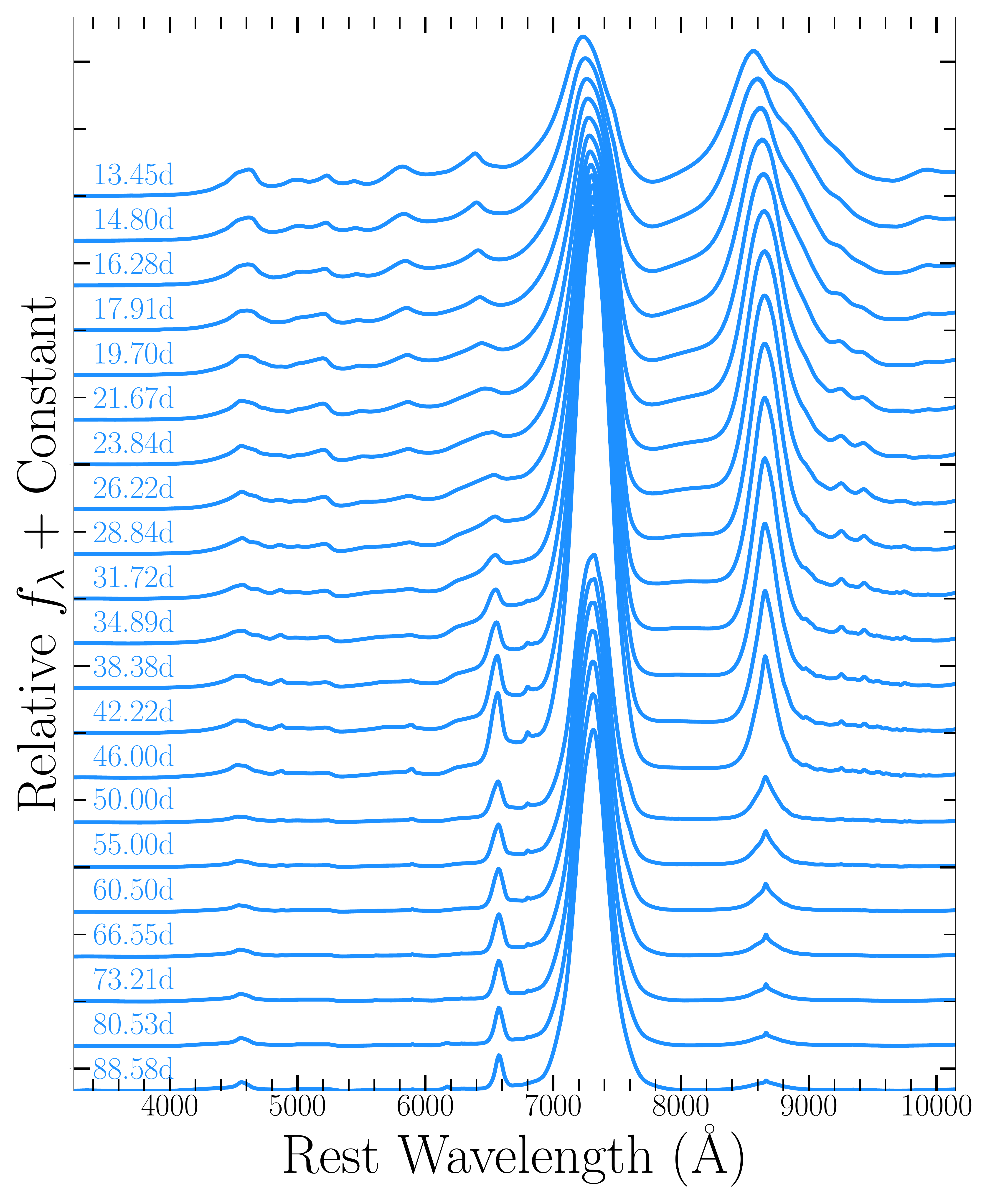}

\caption{Spectral series of fca$_3$  model. Each spectrum is normalized by the mean spectral flux.}
\label{fig:spectra_fca2}
\end{figure*}

\end{document}